\documentclass[prb,amsmath,amssymb]{revtex4-2} 

\usepackage{color}

\usepackage{subcaption}
\definecolor{linkcol}{rgb}{0,0,0.4}
\definecolor{citecol}{rgb}{0.5,0,0}
\usepackage[pdftex]{graphicx}
\usepackage{overpic}

\definecolor{jerem}{rgb}{1, 0, 0}
\definecolor{jerem}{rgb}{1, 0, 0}
\definecolor{jerem2}{rgb}{0, 0, 1}
\definecolor{olivier}{rgb}{0.125, 0.26, 0.07}

\newcommand{\der}[0]{\text{d}}

\newcommand{\ob}[0]{\mathcal{B}}
\newcommand{\xnh}[0]{\mathbf{x_0}}
\newcommand{\uh}[0]{\mathbf{u}}

\newcommand{\bo}[1]{\mathbf{#1}}

\newcommand{\rev}[1]{\textcolor{black}{#1}}

\usepackage{mathtools}
\usepackage{footnote}

\usepackage{makecell}

\usepackage{hyperref}

\hypersetup
{
	bookmarksopen=true,
	pdftitle="Manuscript title",
	pdfauthor="Your name",
	pdfsubject="Manuscript topic in a few words", 
	pdfmenubar=true, 
	pdfhighlight=/O, 
	colorlinks=true, 
	pdfpagemode=UseNone, 
	pdfpagelayout=SinglePage, 
	pdffitwindow=true, 
	linkcolor=linkcol, 
	citecolor=citecol, 
	urlcolor=linkcol 
}

\usepackage{enumerate}
\usepackage{mathtools}
\usepackage{stmaryrd}

\usepackage{enumitem}

\usepackage{epstopdf}

\usepackage{textcomp}
\usepackage{gensymb}
\usepackage{lipsum} 

\def \equi#1{\mathrel{\mathop{\kern 0pt\sim}\limits_{#1}}}

\newcommand{\figwone}[0]{0.22\textwidth}
\newcommand{\figwtwo}[0]{0.9\textwidth}

\begin{document}
\title{\rev{Large volume statistics of first-passage observables of $d$--dimensional Jump Processes}}
\author{J. Klinger}
\affiliation{Centre Sciences des Données, ENS/PSL, 
 45 rue d'Ulm, 75005 Paris, France}
 \author{O. B\'enichou}
\affiliation{Laboratoire de Physique Th\'eorique de la Mati\`ere Condens\'ee, CNRS/Sorbonne Université, 
 4 Place Jussieu, 75005 Paris, France}
 \author{R. Voituriez}
\affiliation{Laboratoire Jean Perrin, CNRS/Sorbonne Université, 
 4 Place Jussieu, 75005 Paris, France}

\begin{abstract}

First-passage observables (FPO) are central to understanding stochastic processes in confined domains, with applications spanning chemical reaction kinetics, foraging behavior, and molecular transport. While extensive analytical results exist for continuous processes, discrete jump processes—crucial for describing empirically observed dynamics—remain largely unexplored in this context. This paper presents a comprehensive framework to systematically evaluate FPO for $d$-dimensional isotropic jump processes, \rev{in the asymptotic regime of large confining domains}, capturing both geometric and dynamical observables. Leveraging the connection between jump processes and their continuous counterparts, we address the limitations of continuous approximations in capturing discrete effects, particularly near absorbing boundaries. Our method unifies FPO calculations across edge and bulk regimes, providing explicit asymptotic expressions for key observables, such as splitting probabilities, \rev{distributions of escape points}, and first-passage time distributions in complex geometries. We illustrate our approach with paradigmatic examples, \rev{including splitting probabilities in two-dimensional eccentric disks and mean exit times for heavy-tailed processes}. These results underscore the broad applicability of our framework to diverse physical systems, offering novel insights into the \rev{complexity of discrete dynamics in bounded geometries.} 

\end{abstract}
\date{\today}

\maketitle

\newpage
\section{Introduction}

\begin{figure}[h!]
	\centering
	\begin{subfigure}{\figwone}
		\centering
		\includegraphics[height=\textwidth]{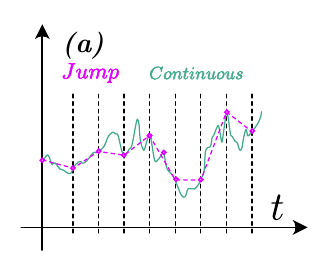}
	\end{subfigure}
	\hspace{8pt}\begin{subfigure}{\figwone}
		\centering
		\includegraphics[height=\figwtwo]{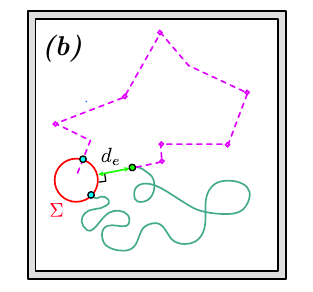}
	\end{subfigure}
	\begin{subfigure}{\figwone}
		\centering
		\includegraphics[height=\figwtwo]{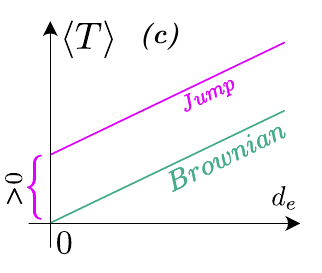}
	\end{subfigure}

	\begin{subfigure}{\figwone}
		\centering
		\includegraphics[height=\figwtwo]{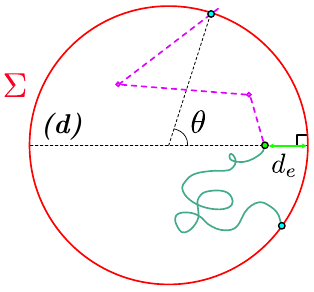}
	\end{subfigure}
	\hspace{-8pt}
	\begin{subfigure}{\figwone}
		\centering
		\includegraphics[height=\figwtwo]{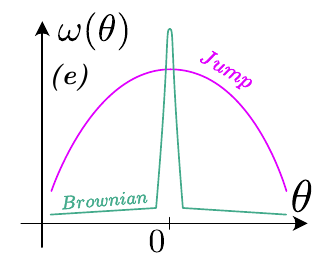}
	\end{subfigure}
	\caption{
		\textbf{(a)} Limited time resolution of continuous dynamics induces empirical realizations of jump processes and strongly impacts the statistics of both geometric and dynamical first passage time observables, across both target-type and exit-type problems. \textbf{(b-c)}  Dynamical observable in a target-type problem - the mean first passage time $\langle T \rangle$ to an absorbing interior target decreases as a function of the source target distance $d_e$. However, the discrete nature of jump processes \rev{allows} for a non-vanishing $\langle T \rangle$ for $d_e=0$, in striking contrast to the continuous behavior. \textbf{(d-e)}  Geometrical observable in an exit-type problem - the distribution $\omega(\theta)$ of the exit angle of a Brownian particle in a confining disk converges to a Dirac as $d_e \to 0$. \rev{A contrario}, the same distribution for jump processes is never singular.
	}
	\label{fig:schematic_1}
\end{figure}

\rev{Stochastic dynamics in finite domains}
are essential models to describe a variety of physical and chemical systems, from (bio)chemical catalytic reactions \cite{RiceBook} to animal behavior in delimited ecosystems \cite{jeanson:2005}. In this context, of prime importance is the quantification of first-passage observables (FPO) associated with either exit events from a  domain through its boundary (see Fig. \ref{fig:schematic_1}d), or first encounters typically with  small target subdomains located in the bulk of the  \rev{bounding} domain (see Fig. \ref{fig:schematic_1}b),  \rev{which often constitute} kinetically limiting steps for various processes of interest.

Numerous and varied illustrations abound for both these exit and target problems. Examples of exit problems include
chemical reaction rates that can be effectively modeled via mean exit times from local minima of complex confining energy landscapes \cite{Arrhenius:1889,kramers_brownian_1940},
allelic fixation probabilities \cite{Wright1931} that derive from splitting probabilities \cite{Feller2:1971} of escaping an interval through one side rather than the other,
or the exploratory behavior of animals that can be quantified by the mean exit time from a given habitat. In another context, portfolio returns are estimated by computing overshoot (also referred to as hitting) distributions of one-dimensional processes upon crossing a fixed level $x$ \cite{Kou:2003}. Target problems can be exemplified at various scales, from the binding kinetics of a molecule with its specific ligand, immune cells looking for infection sites in tissues, to larger scale animals searching for food \cite{RandonFurling:2009}.

Analytical results for  exit and target problems can be broadly categorized into two general classes of observables, depicted on Fig. \ref{fig:schematic_1}.

First, geometrical FPO are associated to the spatial properties of the trajectories of a model particle, and belong to the class of splitting probabilities, defined generically  as the probability to reach first  a given subdomain among a set of absorbing subdomains, irrespectively of the time  involved. Because subdomains can be either  located in the bulk of the confining domain, or be the domain boundary itself, geometrical FPO can either be classified as target or exit problems.  Among important results, the \rev{continuous} one-dimensional splitting probability $\pi^{(c)}_{0,\underline{x}}(x_0)$ of escaping a confining interval $[0,x]$ through $x$ rather than through $0$ has been explicitly computed for both Brownian and $\alpha$-stable processes \cite{Feller2:1971,Majumdar2010,Majumdar:2010,Kyprianou:2022p}. In higher dimensional  geometries, splitting probabilities to small  targets in the bulk of a confining domain have been obtained asymptotically for Brownian processes
\cite{Condamin:2005,Condamin:2008,Pillay:2010,chevalier_first-passage_2011}; in spherical geometries, \rev{the distribution of the position of the particle upon exiting the domain (classically referred to as the harmonic measure in the case of Brownian dynamics) is also known for arbitrary interior starting position for both Brownian and $\alpha$-stable processes \cite{Redner:2001a, Blumenthal1961}}. 

Secondly, dynamical FPO quantify the time taken by the particle to reach a given subdomain (either small and located in the bulk of the confining domain, or  the domain boundary itself).  Mean first-passage times to small targets in the bulk of a confining domain have been obtained asymptotically for Brownian processes \cite{chevalier_first-passage_2011,cheviakov_optimizing_2011}, while  mean first exit times are known explicitly in spherical geometries for both  Brownian and $\alpha$-stable  processes \cite{Redner:2001a,Getoor1961}.  Asymptotic forms of the corresponding complete distributions have also been obtained  \cite{Meyer:2011}
, as well as their generalizations to  general scale invariant processes in the continuous limit \cite{BenichouO.:2010}.

Importantly, essentially all determinations of FPO available so far apply to processes that are continuous in \rev{both} space and  time.  However, in numerous cases the dynamics of a particle (or agent) of interest cannot be described by continuous-time processes, \rev{and its motion is better modeled by discrete-time and continuous-space processes, also known as \textit{jump processes} \cite{VanKampen:1992,Berg:2004, Araujo:2021,Vezzani:2024aa}}. In the one-dimensional setting, a jump process, starting from an initial position $x_0$, is defined by its position $x_n$ at  discrete time $n$, which is given by $x_n=x_{n-1}+\xi_n$,
where the $\xi_i$'s are \textit{iid} random variables with common distribution $p(\ell)$, which will be taken to be symmetric. In turn, the distribution  $p(\ell)$ completely defines the dynamics. The importance of jump processes is twofold. 
On the one hand they are naturally suited to account for empirical trajectories displaying randomly reorienting ballistic motion, whether it be in light scattering experiments \cite{Blanco:2006, Baudouin:2014} or tracking of self-propelled particles \cite{Romanczuk:2012fk, Solon:2015}.
On the other hand, typical experimental data provide  time series that are intrinsically discretized due to sampling \rev{procedures}, independently of the underlying dynamics (see Fig \ref{fig:schematic_1}a). Consequently, any observable that is measured from experimental data relies on discrete time series, and \rev{its statistics} cannot, by definition, be directly inferred from  continuous-time models. As we show below, the discrete nature of jump processes can have dramatic consequences on a broad class of FPO (see Fig \ref{fig:schematic_1}c,e).

Despite this broad relevance, there \rev{are} to date no general results on FPO of  \textit{jump processes in finite domains}, with the exception of the determination of the asymptotic one-dimensional splitting probability $\pi_{0,\underline{x}}(x_0)$ \cite{Klinger2022}  (note that hereafter observables for jump processes will be denoted without \rev{superscript}, while their continuous-time counterpart\rev{s} have a \rev{superscript $(c)$}). This stems from the intrinsic complexity of the integral equations that define FPO of jump processes, even in the simplest setting of a 1--dimensional interval \cite{VanKampen:1992}.

To overcome this technical difficulty, it is natural to analyze the continuous limit of FPO of jump processes. Indeed, in the large $n$ limit, jump processes converge to well known limit continuous processes \cite{Gnedenko:1955, Stone:1967} , which depend only on the small $k$ behavior of the Fourier Transform $\tilde{p}(k)=\int_{-\infty}^{\infty}e^{ik\ell}p(\ell)\der \ell$ of the jump distribution:
\begin{equation}\label{eq:chap4_ft_p}
	\tilde{p}(k)\underset{k\to 0}{=}1-(a_\mu |k|)^\mu + o(k^\mu).
\end{equation}
\noindent Two situations arise: when the \rev{L\'evy} index $\mu$ is equal to 2, the limit process is a Brownian motion with diffusion coefficient $D=a_2^2$; when $\mu\in]0,2[$ the jump process is dubbed heavy-tailed, and converges to an $\alpha$-stable process whose distribution is given in Fourier space by $\tilde{P}(k,n)=\exp\left[-n |a_\mu k|^\mu\right]$. 

However, quite unexpectedly, FPO of confined jump processes in general cannot be deduced directly from the knowledge of the corresponding FPO for the limit continuous process. This is best appreciated by considering the survival probability $q_{[0,x]}(x_0,n)$  defined as the probability for a one-dimensional jump process to remain inside the interval $[0,x]$ during its first $n$ steps. In the large volume limit $x\to \infty$, $q_{[0,x]}(x_0,n)$ \rev{converges to the} semi-infinite survival probability $q(x_0,n)$ that the particle remains positive during its first $n$ steps, which was obtained in \cite{Ivanov1994}. 
Crucially, owing to the discrete nature of the dynamics, the survival probability from $x_0=0$ is strictly positive, as determined by the renowned Sparre-Andersen theorem \cite{Sparre:1954}

\begin{equation}\label{eq:chap4_sparre}
	q(0,n) = \binom{2n}{n}\frac{1}{2^{2n}},
\end{equation}

\noindent in stark contrast with the small $x_0$ behavior of the limit continuous semi-infinite survival probability $q^{(c)}(x_0,n)$ \cite{Majumdar2017}

\begin{equation}\label{eq:chap4_identification_3}
	q^{(c)}(x_0,n)\underset{x_0 \to 0}{\sim}\frac{1}{\sqrt{n}}\frac{a_\mu^{-\frac{\mu}{2}}}{\sqrt{\pi} \Gamma(1+\frac{\mu}{2})} x_0^{\frac{\mu}{2}},
\end{equation}

\noindent which identically vanishes as $x_0\to 0$. Consequently, the limit FPO results are clearly ill-fitted to describe the complete range of initial $x_0$ positions, and  discrete effects dramatically impact FPO at all times, even in the large volume limit. This limitation is in fact very general and holds for all FPO also in higher dimensions, as soon as the initial position of the \rev{jump process} belongs (or is close) to an absorbing boundary, \textit{i.e.}, for all return-type problems as exemplified in Fig. \ref{fig:schematic_1}b-e. This class of return-type problems covers both target and exit problems, and geometrical or dynamical FPO; they  have proved to have a broad range of applications, from molecular transport \cite{Anikeenko:2009aa}, light scattering \cite{Savo:2017}, animal behavior \cite{Blanco:2003a} or even robotics \cite{Hidalgo-Caballero:2024aa}. In all such cases, theoretical predictions based on continuous limits yield vanishing FPO, and are thus {\it irrelevant}. 
Determining FPO of jump processes in \rev{finite} domains, in particular for return-type problems,  thus constitutes  an open theoretical challenge, with  a broad range of applications in physics and beyond.

In this article, we develop a general framework to systematically evaluate FPO statistics of $d$-dimensional 
isotropic jump processes, in the large 
volume limit, and for all initial positions $x_0$. In particular we properly characterize the regime where $x_0$ is close to the absorbing boundary, for which  continuous limits  fail to capture discrete effects. In fact, we show that FPO statistics can be obtained explicitly from the combined knowledge of their continuous counterpart \textit{and} the one-dimensional splitting probability of an auxiliary jump process. In turn, we illustrate our framework by deriving asymptotic universal formulas for paradigmatic examples of geometrical and dynamical observables, in one and higher dimensions.

More precisely, we focus in Section II on hitting distributions of one-dimensional jump processes, and show how the splitting probability allows us to capture exactly the non-vanishing behavior of FPO \rev{in the $d_e\to 0$ limit}, where  $d_e$ denotes the distance of $x_0$ to the absorbing boundary of the domain.

Building upon this cornerstone result, we present and discuss the general method in Section III, which is  \rev{asymptotically valid for arbitrary isotropic processes evolving in finite domains with fully absorbing boundaries}.
We focus in Section IV on illustrating the method with relevant examples of  geometrical observables, among which hitting distributions in spherical geometries and splitting probabilities in disks with inclusions. Section V is dedicated to dynamical observables, where we compute mean exit times \rev{in} spherical geometries \rev{and} complete exit time distributions. \rev{Finally, in Section VI we extend our framework to confining domains with a reflecting boundary, focusing on processes with $\mu=2$, for which reflecting boundary conditions are unambiguously defined both for the discrete-time jump process and its continuous-time Brownian limit; as an illustration, we compute the first-passage time distribution to an interior absorbing target.}

\section{Geometrical First-passage observables of $1d$ jump processes}\label{seq:2}

As announced in the introduction, the only general result for FPO of jump processes available in the literature concerns the splitting probability $\pi_{0,\underline{x}}(x_0)$ in the large 
volume limit $x\to \infty$. In the following, we show that $\pi_{0,\underline{x}}(x_0)$ plays a pivotal role in the determination of general FPO behavior, and thus remind its expression for consistency in subsection A below. As a first application, we derive in subsection B the hitting distribution $f_{[0,x]}(y|x_0)$ of the landing position $y>x$ upon first exit to the right of $[0,x]$ for all initial positions $x_0$.

\subsection{Splitting Probabilities}

The splitting probability $\pi^{(c)}_{0,\underline{x}}(x_0)$ of continuous $\alpha$-stable processes is known \cite{Blumenthal1961} for all $\mu\in]0,2]$ and given by 

\begin{equation}\label{eq:chap4_identification_1}
	\pi^{(c)}_{0,\underline{x}}(x_0) = \frac{\Gamma(\mu)}{\Gamma^2(\frac{\mu}{2})}\int_0^{\frac{x_0}{x}}\left[u(1-u)\right]^{\mu/2-1} \text{d}u.
\end{equation}
Note that the case $\mu=2$ is contained in that formula, and simply yields the Brownian result \cite{Redner:2001a} $\pi^{(c)}_{0,\underline{x}}(x_0)=x_0/x$.
\noindent \rev{We focus first on the regime of bulk \rev{initial} condition, defined as $a_\mu\ll x_0 \ll x$. On the one hand, the large volume limit hypothesis $x_0 \ll x$ allows the asymptotic approximation  $\pi^{(c)}_{0,\underline{x}}(x_0)\underset{x_0 \ll x}{\sim} \frac{2\Gamma(\mu)}{\mu \Gamma^2(\mu/2)}(x_0/x)^{\mu/2}$. On the other hand, the bulk initial condition $a_\mu\ll x_0$ is the key ingredient to approximate the discrete splitting probability by its continuous counterpart. As a result, the small $x_0$ behavior of equation \eqref{eq:chap4_identification_1} accurately describes the splitting probability of jump processes, such that}
 
\begin{equation}\label{eq:chap4_identification_2}
	\pi_{0,\underline{x}}(x_0)\underset{a_\mu\ll x_0\ll x}{\sim}\frac{2\Gamma(\mu)}{\mu\Gamma^2(\frac{\mu}{2})}\left(\frac{x_0}{x}\right)^{\frac{\mu}{2}}.
\end{equation}

\noindent In other words, when starting in the bulk of the domain, the jump process is correctly described by its continuous limit. However, it is clear that \rev{in the regime of 
edge initial conditions, defined as $0\leq x_0 \lesssim a_\mu$,} equation  \eqref{eq:chap4_identification_2} does not capture the inherent discrete effects, and would incorrectly yield a vanishing splitting probability for $x_0\to 0$, whereas the splitting probability from $x_0=0$ should be strictly positive since $\pi_{0,\underline{x}}(0)>\int_{x}^{\infty}p(\ell)\der \ell > 0$. In fact, the exact asymptotic splitting probability is given by

\begin{equation}\label{eq:chap4_split_main}
	\begin{alignedat}{1}
		&\underset{x\rightarrow\infty}{\lim}\left[\frac{\pi_{0,\underline{x}}(x_0)}{A_\mu(x)} \right]=\frac{1}{\sqrt{\pi}}+V(x_0)\\
		&A_\mu(x)=\Bigg(\frac{a_\mu}{x}\Bigg)^{\mu/2} 2^{\mu-1}\Gamma\left(\frac{1+\mu}{2}\right),
	\end{alignedat}
\end{equation}
\rev{\noindent where $V(x_0)$ is a $p(\ell)$ dependent function \cite{Majumdar2017} satisfying $V(0)=0$, and \rev{explicitly} given by its Laplace transform \cite{Majumdar2017,Ivanov1994}
\begin{equation}\label{eq:chap4_survival_5}
	\int_0^\infty e^{-s x_0}V(x_0)\der x_0 =\frac{1}{s \sqrt{\pi}}\left(\exp\left[-\frac{s}{\pi}\int_0^\infty \frac{\der k}{s^2+k^2}\ln(1-\tilde{p}(k))\right]-1\right).
\end{equation}
}
\rev{Since $V(0)=0$, it is found that the splitting probability from 0 takes the following universal form}

\begin{equation}\label{eq:ch4_spl_0}
	\pi_{0,\underline{x}}(0)\underset{x\rightarrow\infty}{\sim}\frac{2^{\mu-1}}{\sqrt{\pi}}\Gamma\left(\frac{1+\mu}{2}\right)\Bigg(\frac{a_\mu}{x}
	\Bigg)^{\frac{\mu}{2}},
\end{equation}

\noindent which is non-vanishing for all $x$ values, and depends only on the \rev{L\'evy} index $\mu$ and characteristic lengthscale $a_\mu$ of the jump process.

\subsection{Hitting distribution after exit}

For heavy tailed jump processes with $\mu\in]0,2[$, significant overshoots occur upon first crossing of the fixed level $x$, even in the large volume limit. As a first novel result, we derive the hitting distribution $f_{[0,x]}(y|x_0)$ of the landing position $y>x$ upon first exit of $[0,x]$ through $x$ for all initial positions $x_0$, including the edge regime $x_0\lesssim a_\mu$.

\begin{figure}
\centering
\includegraphics[width=0.6\textwidth]{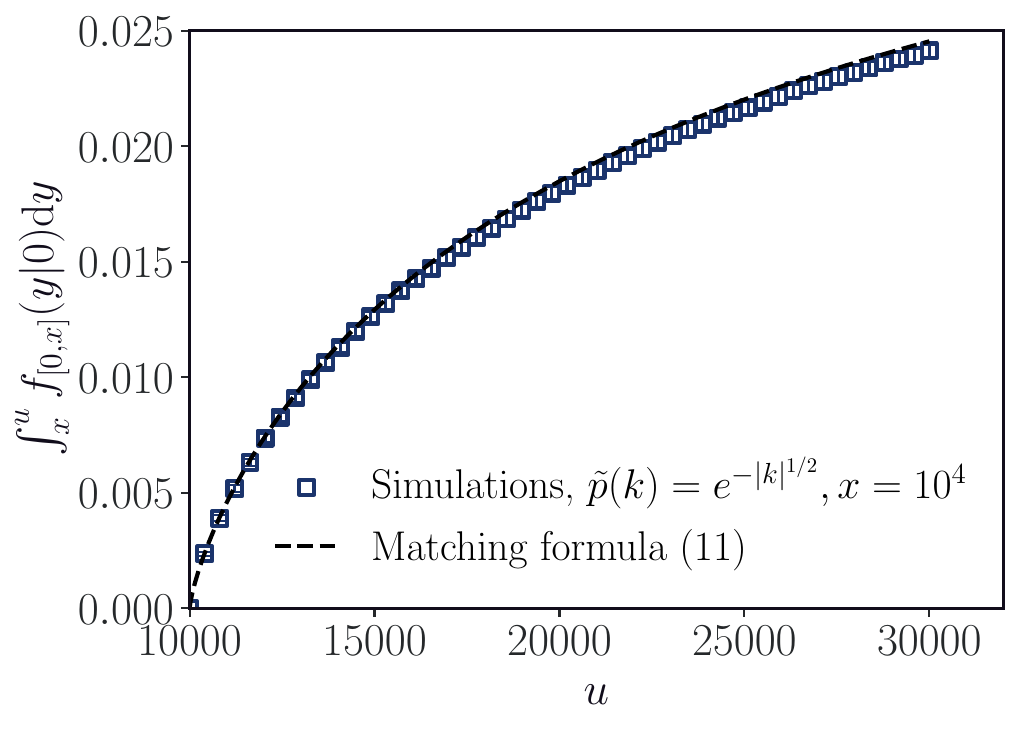}
\caption{
\rev{
Cumulative overshoot distribution for a heavy-tailed \rev{L\'evy} jump process with $\mu=1/2$ starting from $x_0=0$. The dashed line corresponds to the numerical integration of equation \eqref{eq:ch7_hitting_4}. Empirical counts are averaged over $10^6$ particles evolved until the interval is escaped.
Empirical and theoretical results agree perfectly over a broad range of overshoot distances. Of most importance is the fact that the prefactor of the cumulative distribution is perfectly captured. 
We note the slight discrepancy observed at larger $u$ values but attribute it to another type of discretization effect beyond the scope of this work, namely convergence of the statistics of conditional trajectories associated to very large overshoots.
}
}
\label{fig:hitting}
\end{figure}

To that end, we consider the jump process conditioned to cross $x$ before $0$. For such conditioned processes and $y>x$, the  hitting distribution conditioned \rev{on right exit events} is  given by $f_{[0,x]}(y|x_0,{\rm right})=f_{[0,x]}(y|x_0)/\pi_{0,\underline{x}}(x_0)$. 
\rev{
Next, we make use of the large times convergence of the conditioned jump process towards its conditioned continuous limit \cite{Caravenna:2008} to relate both conditional hitting distributions: 
\begin{equation}\label{eq:ch7_hitting_1}
	\begin{alignedat}{1}
		&\frac{f_{[0,x]}(y|x_0)}{\pi_{0,\underline{x}}(x_0)}\underset{x\to\infty}{\sim}\frac{f^{(c)}_{[0,x]}(y|x_0)}{\pi^{(c)}_{0,\underline{x}}(x_0)}\equiv g(y,x),
	\end{alignedat}
\end{equation}
where we additionally use the fact that for $\alpha$-stable processes, the conditional hitting distribution denoted $g$ is asymptotically $x_0$ independent.
Crucially, since equation \eqref{eq:ch7_hitting_1} is valid for all $x_0$ values and the right hand side is $x_0$ independent, we obtain that the entire $x_0$ dependence of the discrete hitting distribution $f_{[0,x]}(y|x_0)$ is carried by the splitting probability $\pi_{0,\underline{x}}(x_0)$ alone. We emphasize that this property is a direct consequence of the conditional distribution $f^{(c)} / \pi^{(c)}$ being asymptotically independent of the initial position $u$.
The determination of the hitting distribution thus reduces to computing its continuous counterpart, since splitting probabilities are known both in the discrete and continuous paradigms. Making use of exact expressions for hitting distributions of $\alpha$-stable processes given in \cite{Blumenthal1961}
}

\begin{equation}\label{eq:ch7_hitting_2}
	\centering
	\begin{alignedat}{1}
		&f^{(c)}_{[0,x]}(y|x_0)=f^{(c)}_{[-1,1]}\left(\frac{2 y}{x}-1 | \frac{2 x_0}{x}-1\right)\frac{2}{x},\\
		&f^{(c)}_{[-1,1]}(y|x_0)=\frac{1}{\pi}\sin\left(\frac{\pi \mu}{2}\right)\left(\frac{|1-|x_0|^2|}{|1-|y|^2|}\right)^{\mu/2}|x_0-y|^{-1},
	\end{alignedat}
\end{equation}

\noindent we obtain the general asymptotic hitting distribution for heavy-tailed jump processes: 

\begin{equation}\label{eq:ch7_hitting_4}
	\frac{f_{[0,x]}(y|x_0)}{1+V(x_0)}\underset{\substack{x\to \infty \\ y>x}}{\sim}\frac{2^\mu \Gamma(1+\frac{\mu}{2})\sin(\frac{\pi \mu}{2})}{\pi\left[\left(\frac{2y}{x}-1\right)^2-1\right]^{\frac{\mu}{2}}}\frac{1}{y}\left[\frac{a_\mu}{x}\right]^{\frac{\mu}{2}},
\end{equation} 

\noindent where $V(x_0)$ is defined in equation \eqref{eq:chap4_survival_5}. Equation \eqref{eq:ch7_hitting_4} is  illustrated in Fig. \ref{fig:hitting} for the specific heavy tailed distribution $\tilde{p}(k)=e^{-|k|^{1/2}}$ and initial position  $x_0=0$.

In computing the hitting distribution, we have shown that the one-dimensional splitting probability and the conditional limit continuous process play an essential role to overcome the challenge posed by the vanishing small $x_0$ behavior of continuous FPO. In the following, we expand upon these ideas and propose a comprehensive framework to compute \rev{asymptotic statistics of} general jump process FPO of both geometrical and dynamical types, in one or more dimensions, and adaptable to diverse boundary conditions. 

\section{General matching method for $d$-dimensional jump processes}\label{seq:3}
Hereafter we consider  isotropic $d$-dimensional jump processes, whose single jump distribution $p(\boldsymbol{\ell})$, $\boldsymbol{\ell} \in \mathbb{R}^d$ depends only on the norm $|\boldsymbol{\ell}|\equiv\ell$. 
\rev{As in the one-dimensional case (see Equation \eqref{eq:chap4_ft_p}), the large $\boldsymbol{\ell}$ behavior of the jump distribution is equivalently given by the small $\boldsymbol{k}$ behavior of the Fourier transform $\tilde{p}(\boldsymbol{k})$ of the jump distribution}
:

\begin{equation}\label{eq:ch7_ddim_jp}
	\tilde{p}(\boldsymbol{k}) = \int_{\mathbb{R}^d}e^{i \boldsymbol{k}\cdot \boldsymbol{\ell} }p(\boldsymbol{\ell})\text{d}\boldsymbol{\ell}\underset{|\boldsymbol{k}|\to 0}{=}1-|a_\mu \boldsymbol{k}|^{\mu}+o(|\boldsymbol{k}|^{\mu}).
\end{equation}

\noindent Note that equation \eqref{eq:ch7_ddim_jp} also defines the limit distribution of the jump process: $d$-dimensional isotropic $\alpha$-stable for $\mu<2$ and else Brownian with $D=a_2^2$.

\begin{figure}[h!]
	\centering
	\begin{overpic}[height=0.4\textwidth]{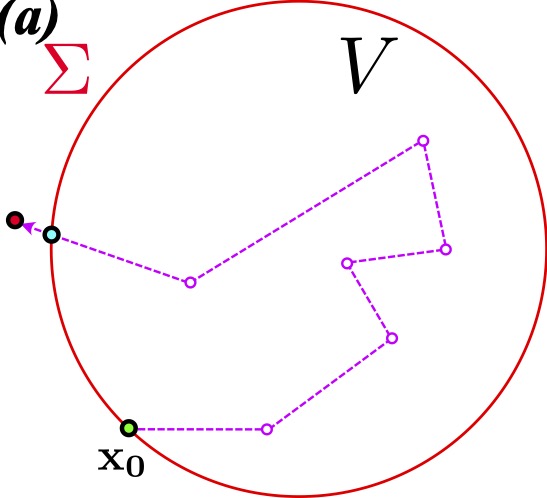}
		\put(0,83){\color{white}\rule{14\unitlength}{10\unitlength}}
	\end{overpic}
	\caption{Two-dimensional isotropic jump process in a fully absorbing  disk, starting from $\xnh$. Here the absorbing boundary $\Sigma$ corresponds to the complete disk boundary, and the process is stopped as soon as it strictly exits the disk. \rev{Recalling $n^*$ is the exit step, the escape point is here defined as the intersecting point  (blue on the figure) between the absorbing disk and the segment joining $\mathbf{x}_{n^{*}-1}$ and $\mathbf{x}_{n^{*}}$.}
	}
	\label{fig:chap7_schem_1}
\end{figure}

Starting from $\xnh$, the particle evolves in a \rev{finite, fully absorbing} domain of volume $V$ with absorbing boundary $\Sigma$, as depicted in Fig. \ref{fig:chap7_schem_1}. 
\rev{We consider in the following arbitrary first-passage observables $\ob(\xnh)$, evaluated upon first crossing of the boundary $\Sigma$. That is, denoting $n^*= \inf\{n; \mathbf{x}_n \in \bar{V}\}$ the exit step, $\ob$ is any observable that can be evaluated from the position pair $(\mathbf{x}_{n^{*}-1} , \mathbf{x}_{n^{*}})$.}
To be more explicit, the distribution of the random variable $\ob(\xnh)$ covers the example of the one-dimensional splitting probability $\pi_{0,\underline{x}}(x_0)$ to escape the interval $[0,x]$ through $x$; for $d=2$ it can be exemplified by  \rev{the distribution of escape points (probability density to escape the domain through a given point of its boundary, see Fig. \ref{fig:chap7_schem_1})}, or the splitting probability to reach an interior absorbing target before the (also absorbing) domain boundary, as illustrated in Section~\ref{seq:4}\rev{.A} below with the example of an eccentric annulus.
\noindent In the following, we assume that the domain $V$ has a unique characteristic lengthscale $R$, and define the large volume limit as  $R\gg a_\mu$. Importantly, because of the discrete nature of jump processes,  $\ob(\xnh)$ is well defined for all initial positions $\xnh$ such that $\xnh$ belongs to the interior or the boundary of the domain. In the following, we investigate separately the case where $\xnh$ is close to or far from the absorbing boundary.

\begin{figure}[h!]
	\centering
	\includegraphics[width=0.4\textwidth]{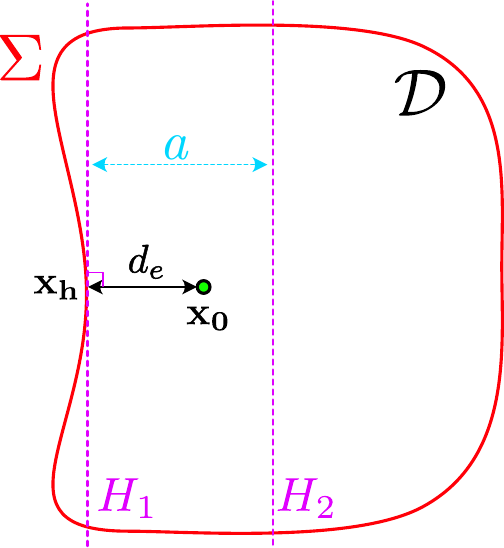}
	\caption{ 
	\textbf{Schematics of the matching method.} The full boundary $\Sigma$ is absorbing, and the starting position $\xnh$ is close to $\Sigma$. The two delimiting hyperplanes $H_1$ and $H_2$ are parallel, and $H_1$ is tangent to $\Sigma$. Importantly, we consider the edge regime $d_e\lesssim a_\mu \ll a\ll R$. In this regime, \rev{the statistical properties of trajectories conditioned to cross $H_2$ before $H_1$ are asymptotically equivalent to those of the continuous limit process.}
	}
	\label{fig:ch7_sandwitch}
\end{figure}

To delineate these hereafter called  \textit{edge} and \textit{bulk} regimes in high-dimensional domains, we introduce the orthogonal projection $\bo{x_h}$ of $\xnh$ on the absorbing surface $\Sigma$, and the corresponding source-surface distance $d_e=|\bo{x_h}-\xnh|$, as illustrated on Fig. \ref{fig:ch7_sandwitch}. In turn, we define the bulk initial conditions as the regime $d_e\gg a_\mu$ (and thus $R\gg a_\mu$). Following the one-dimensional treatment, the \rev{corresponding statistical} bulk behavior of $\ob(\xnh)$ is simply given by

\begin{equation}\label{eq:ch7_stitching_1}
	\ob(\xnh)\underset{a_\mu\ll d_e \ll R}{\sim} \ob^{(c)}(\xnh),
\end{equation}
where $\ob^{(c)}(\xnh)$ corresponds to the same FPO, evaluated for the continuous limit process. We emphasize that \rev{the two key hypothesis for equation \eqref{eq:ch7_stitching_1} to be useful are that $a_\mu$ is small with respect to the spatial extension of $\Sigma$, and that the continuous statistics of $\ob^{(c)}(\xnh)$ can be determined explicitly.  
Note however that these conditions are met for a large number of first-passage observables, as discussed above. }

In the edge regime $d_e\ll a_\mu$, and in particular in the specific case $\xnh \in \Sigma$, equation \eqref{eq:ch7_stitching_1} is not valid because all continuous trajectories almost surely exit the domain through $\xnh$  immediately. To circumvent this difficulty we employ a conditioning procedure akin to the one applied in the one-dimensional case. Specifically, we introduce two parallel hyperplanes $H_1$ and $H_2$ distant from an arbitrary distance $a$, such that $H_1$ is tangent to $\Sigma$ at $\mathbf{x_h}$ (see Fig. \ref{fig:ch7_sandwitch}).

\rev{
By partitioning trajectories contributing to the distribution of $\ob(\xnh)$ according to whether $H_1, H_2$ or $\Sigma$ is first crossed we obtain the following exact expression:
\begin{equation}\label{eq:ch7_stitching_2}
	\hspace{-20pt}
	\begin{alignedat}{1}
		\mathbb{P}(\ob(\xnh)=b)=&\pi_{H_1,\underline{H_2}, \Sigma}(\xnh)\mathbb{P}[\ob(\xnh)=b | \text{$H_2$ crossed before $H_1, \Sigma$}]\\
		+&\pi_{\underline{H_1},H_2, \Sigma}(\xnh)\mathbb{P}[\ob(\xnh)=b |\text{$H_1$ crossed before $H_2, \Sigma$}]\\
		+&\pi_{H_1,H_2, \underline{\Sigma}}(\xnh)\mathbb{P}[\ob(\xnh)=b |\text{$\Sigma$ crossed before $H_1,H_2$}],
	\end{alignedat}
\end{equation}
where $\pi_{H_1,\underline{H_2}, \Sigma}(\xnh)$ is the three-way splitting probability that the particle crosses $H_2$ before $H_1$ or $\Sigma$.
Equation \eqref{eq:ch7_stitching_2} is an exhaustive partition of trajectories, coarsely exemplified in Fig. \ref{fig:ch7_sandwitch}; the particle crosses $H_2$ by heading rightwards, $H_1$ by heading leftwards or $\Sigma$ if it assumes a mostly vertical motion.
To proceed further we aim to characterize the large volume behavior of equation \eqref{eq:ch7_stitching_2} within the boundary layer asymptotic regime $a_\mu \ll a \ll R$. Note that the specific form of the $a\equiv a(R)$ is not relevant as long as $a(R)=o(R)$. The analysis is carried out in two steps. 
}

\rev{
First, we show that the terms associated with a first crossing of $H_1$ or $\Sigma$ generally represent, for two different reasons, subdominant contributions to the statistics of $\ob$. 
\begin{itemize}
	\item Consider trajectories crossing $\Sigma$ first, that is staying in the hyperplane sandwich until the domain is exited. Since $a\ll R$, the three way splitting probability is asymptotically exponentially suppressed, so that $\pi_{H_1,H_2, \underline{\Sigma}}$ can be neglected. Furthermore, this limit also yields the equivalence of the relevant three way splitting probability to the hyperplane splitting probability $\pi_{H_1, \underline{H_2},\Sigma} \underset{R \to \infty}{\sim}\pi_{H_1, \underline{H_2}}$.
	\item Consider next trajectories crossing $H_1$ first. In the large volume limit $a_\mu \ll R$ we argue that such trajectories are typically absorbed through $\Sigma$ in the vicinity of $\mathbf{x_h}$. 
	Depending on the nature of $\ob$, such trajectories constitute a negligible contribution to the statistics. On the one hand, for geometrical observables, the portion of $\Sigma$ reachable by trajectories that cross $H_1$ before $H_2$ shrinks to zero as the system size grows, so their contribution to the statistics of $\ob$ vanishes. For dynamical observables on the other hand, such trajectories are instead absorbed on a microscopic timescale subdominant compared to the typical first-passage timescale $t_R\sim R^\mu$.
\end{itemize}
We emphasize that these steps strongly rely on locally approximating the absorbing surface $\Sigma$ by the tangent hyperplane $H_1$ on a sufficiently large spatial scale, which requires $\Sigma$ to be smooth in the vicinity of $\mathbf{x_h}$, so that the radius of curvature $R_c$ of $\Sigma$ at $\mathbf{x_h}$ is well defined and satisfies $a_\mu \ll a(R) \ll R_c $. 
This condition, together with the single-lengthscale hypothesis on the confining domain $V$ introduced above, typically excludes configurations in which another portion of $\Sigma$, for instance a second, nearby absorbing target, lies within a distance comparable to $a(R)$ of $\mathbf{x_h}$. This excludes, in particular, quasi-one-dimensional or very wobbly domains.
Within these assumptions the exact partition \eqref{eq:ch7_stitching_2} reduces to
\begin{equation}\label{eq:ch7_stitching_2_bis}
		\mathbb{P}(\ob(\xnh)=b)\underset{a_\mu, d_e \ll a \ll R }{\sim} \pi_{H_1,\underline{H_2}}(\xnh)\mathbb{P}[\ob(\xnh)=b | \text{$H_2$ crossed before $H_1$}]
\end{equation}
In turn, we rewrite the hyperplane splitting probability in terms of the one-dimensional splitting probability $\pi^\perp_{0,\underline{a}}(d_e)$ that the jump process projected upon a vector $\mathbf{n}_{H_1}$ normal to $H_1$ reaches $a$ before 0. Importantly, the distribution of the projected process is obtained from that of the isotropic one as $p_\perp(\ell)=\int_{\mathbb{R}^d} \der \boldsymbol{\ell}p(\boldsymbol{\ell})\delta(\boldsymbol{\ell}\rev{\cdot}\mathbf{n}_{H_1}-\ell)$.
Furthermore, as shown in Appendix~\ref{app:projected}, its large $\ell$ decay is identical to that of $p(\boldsymbol{\ell})$ and in the small $k$ Fourier formalism, one has $1-\tilde{p}_\perp(k)\underset{k\to 0}{\sim}(a_\mu |k|)^\mu$ with $\mu, a_\mu$ unchanged.
Consequently, equation \eqref{eq:ch7_stitching_2} is recast as:
\begin{equation}\label{eq:ch7_stitching_3}
		\mathbb{P}(\ob(\xnh)=b)\underset{a_\mu, d_e \ll a \ll R }{\sim} \pi^\perp_{0,\underline{a}}(d_e)\mathbb{P}[\ob(\xnh)=b | \pi_{H_1,\underline{H_2}}(\xnh)].
\end{equation}
}

\rev{We next examine the statistics of the jump process conditioned to cross $H_2$ before $H_1$. Since $a_\mu \ll a$, we use convergence towards the conditional limit process:
\begin{equation}\label{eq:ch7_stitching_4}
	\mathbb{P}[\ob(\xnh)=b | \pi_{H_1,\underline{H_2}}(\xnh)]\underset{a_\mu \ll a \ll R}{\sim}\mathbb{P}[\ob^{(c)}(\xnh)=b | \pi^{(c)}_{H_1,\underline{H_2}}(\xnh)].
\end{equation}
By translation invariance of the flat sandwich geometry along $H_1$, the continuous hyperplane splitting probability reduces to its one-dimensional counterpart, $\pi^{(c)}_{H_1,\underline{H_2}}(\xnh)=\pi^{(c)}_{0,\underline{a}}(d_e)$, so that Eq.~\eqref{eq:ch7_stitching_4} directly relates the discrete quantity of interest to its continuous counterpart evaluated at the same starting point $\xnh$. 
This continuous quantity is further simplified using the self-similarity of the continuous limit process: the ratio $\mathbb{P}[\ob^{(c)}(\mathbf{u})=b]/\pi^{(c)}_{0,\underline{a}}(\mathbf{u})$ can only depend on $\mathbf{u}$ and $a$ through the dimensionless combination $|\mathbf{u}-\mathbf{x_h}|/a$, so that evaluating it at the fixed point $\xnh$ in the boundary-layer limit $a\to\infty$ is equivalent to evaluating it in the limit $\mathbf{u}\to\mathbf{x_h}$ at fixed $a$. As a result, we obtain the matching formula
\begin{equation}\label{eq:ch7_main_result}
	\left[\frac{\mathbb{P}(\ob(\xnh)=b)}{\pi^\perp_{0,\underline{a}}(d_e)}\right]
	\underset{d_e\lesssim a_\mu \ll a \ll R}{\sim}
	\left[\frac{\mathbb{P}(\ob^{(c)}(\xnh)=b)}{\pi^{(c)}_{0,\underline{a}}(d_e)}\right]
	\underset{\substack{R \to \infty \\ \mathbf{u} \to \mathbf{x_h}}}{\sim} \left[\frac{\mathbb{P}(\ob^{(c)}(\mathbf{u})=b)}{\pi^{(c)}_{0,\underline{a}}(\mathbf{u})}\right],
\end{equation}
where the first equivalence is the discrete-to-continuous convergence of Eq.~\eqref{eq:ch7_stitching_4}, and the second is the scaling argument above, with the limit $\mathbf{u}\to\mathbf{x_h}$ taken perpendicularly to $H_1$. Crucially, both ends of this equivalence chain are quantities we can actually compute: the discrete projected splitting probability $\pi^\perp_{0,\underline{a}}(d_e)$ on the left, and a continuous object on the right, whose $\mathbf{u} \to \mathbf{x_h}$ limit exists as a consequence of an eigenfunction decomposition of continuous first-passage observable distributions, discussed in Appendix~\ref{app:eigen}.
}

\rev{
Finally, we emphasize that we have reduced the asymptotic evaluation of FPO of jump processes to the computation of a single one-dimensional quantity, the projected splitting probability $\pi^\perp_{0,\underline{a}}(d_e)$, which fully encompasses the discrete features of the process,
going beyond the classical continuous approach \eqref{eq:ch7_stitching_1}. This general matching method gives access to a variety of new important results for representative examples of confined jump processes, spanning numerous observables and geometries.\\
\indent We remark that the above given derivation remains heuristic in the sense that it does not meet mathematical formalism, and we discussed its scope of validity and various failure modes above. However, it proves sufficient to understand the statistical behavior of jump processes in a  variety of situations highlighted
in the next two sections, where we demonstrate the versatility and flexibility of this  method by computing geometrical and dynamical first-passage observables associated to both exit and  target-type problems in one or more dimensions, with potential relevance to the description of various real physical systems.
}

\section{$d$-dimensional Geometrical First-passage observables}\label{seq:4}

Geometrical FPO capture the spatial properties of stochastic trajectories upon first encounter with the absorbing surface $\Sigma$ of the domain, and are key to quantifying the likelihood of spatially dependent events, such as catalytic chemical reactions \cite{RiceBook}.
As an illustration of our general  matching method, we compute two novel non-trivial geometrical FPO of jump processes with $\mu=2$ in disks with absorbing boundaries: the splitting probability to the boundary in the presence of an absorbing off-centered inclusion, and \rev{the distribution of escape angles on the boundary} (see Fig. \ref{fig:ch7_geom_disk},\ref{fig:ch7_geom_disk_1}).

\subsection{Splitting probability : off-centered inclusion}

\begin{figure}[h!]
	\centering
	\includegraphics[width=0.6\textwidth]{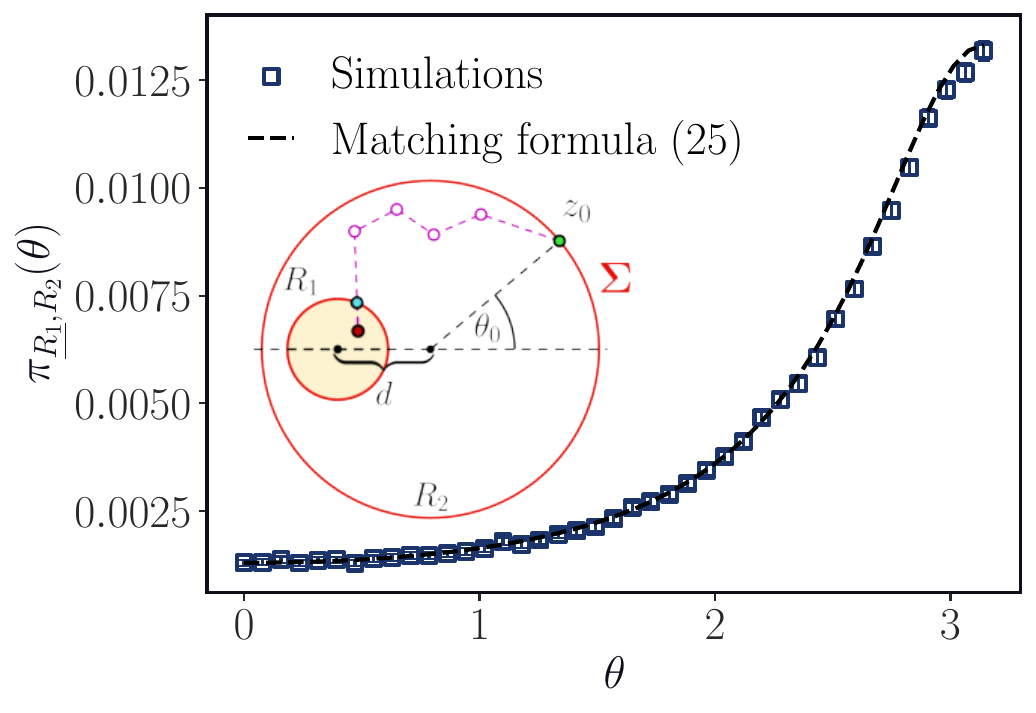}
	\caption{Splitting probability of a \rev{Run-and-Tumble particle (RTP)} to the inner inclusion of an eccentric annulus, starting from the outer rim at position $z_0=R_2e^{i\theta_0}$, with $R_2 = 500$, $R_1=300$ and \rev{inter center distance $d = 150$}.
	\rev{Here, the process is stopped as soon as it ``flies over" the absorbing boundary $\Sigma$. Note that in the large $R_1$,$R_2$ regime, we neglect the small but non zero probability for the process to overshoot the inner inclusion (i.e. there exists a segment linking two successive particle positions and intersecting the inner inclusion boundary twice)}. \rev{Empirical splitting probabilities are estimated from $10^6$ particles per angle $\theta$, evolved until either the inner or outer disk is reached.} The theoretical dashed line is given by equation \eqref{eq:ch7_split_ecc_5}.}
	\label{fig:ch7_geom_disk}
\end{figure}

We first consider the case of a two-dimensional jump process with $\mu=2$ and lengthscale $a_2$, evolving between
\rev{
two eccentric disks of respective radii $R_i$ and center distance $d$ (see inset of Fig. \ref{fig:ch7_geom_disk} for a schematic of the situation), and compute the probability to cross the inner rim first.
}
Note that this geometrical FPO is a typical observable associated to target-type problems - for example catalytic reactions occur only if the catalytic site is found before the ligand exits the reaction domain \cite{Klinger:2022vc}.

Without loss of generality, we consider that the particle starts on the outer disk of radius $R_2$, at position $z_0=R_2 e^{i \theta}$, and denote $\pi_{\underline{R_1},R_2}(\theta)$ the  splitting probability to  cross the inner disk before the outer one. Direct application of the matching method \eqref{eq:ch7_main_result} yields:

\begin{equation}\label{eq:ch7_split_ecc_1}
\underset{R_2 \geq R_1\gg a \gg a_2	}{	\pi_{\underline{R_1},R_2}(\theta) \sim \pi^\perp_{0,\underline{a}}(0)}\underset{u \to z_0}{\lim}\left[\frac{\pi^{(c)}_{\underline{R_1},R_2}(u)}{\pi^{(c)}_{0,\underline{a}}(u)}\right]
\end{equation}
with $u\in\mathbb{C}$. Importantly, the limit $u\to z_0$ has to be taken perpendicularly to the $H_1$ plane, \textit{i.e.}, with fixed $\theta$. Rewriting $u = R_2(1-\epsilon) e^{i\theta}$ with $\epsilon\geq 0$, equation \eqref{eq:ch7_split_ecc_1} is recast as

\begin{equation}\label{eq:ch7_split_ecc_2}
	\pi_{\underline{R_1},R_2}(\theta)\underset{R_2 \geq R_1\gg a \gg a_2	}{\sim}\pi^\perp_{0,\underline{a}}(0)\underset{\epsilon \to 0}{\lim}\left[\frac{\pi^{(c)}_{\underline{R_1},R_2}(R_2(1-\epsilon) e^{i\theta})}{\pi^{(c)}_{0,\underline{a}}(R_2(1-\epsilon) e^{i\theta})}\right].
\end{equation}

\noindent In the small $\epsilon$ limit, the continuous hyperplane splitting probability and projected splitting probability are given by

\begin{equation}\label{eq:ch7_split_ecc_3_1}
	\begin{alignedat}{1}
		\pi^\perp_{0,\underline{a}}(0)\underset{a\gg a_2}{\sim}\frac{a_2}{a},\hspace{10pt}\pi^{(c)}_{0,\underline{a}}((R_2-\epsilon) e^{i\theta})\underset{\epsilon \to 0}{\sim}\frac{\epsilon}{a}
	\end{alignedat}
\end{equation}
such that
\begin{equation}\label{eq:ch7_split_ecc_3_1_1}
	\frac{\pi^\perp_{0,\underline{a}}(0)}{\pi^{(c)}_{0,\underline{a}}((R_2-\epsilon) e^{i\theta})}\underset{\epsilon \to 0}{\sim} \frac{a_2}{\epsilon}
\end{equation}
Next, we compute in \rev{Appendix~\ref{app:eccentric}} the exact eccentric Brownian splitting probability, and show in particular that for small $\epsilon$,  $\pi^{(c)}_{\underline{R_1},R_2}((R_2-\epsilon ) e^{i\theta})$ vanishes linearly

\begin{equation}\label{eq:ch7_split_ecc_3_2}
	\pi^{(c)}_{\underline{R_1},R_2}((R_2-\epsilon) e^{i\theta})\underset{\epsilon \to 0}{\sim} \epsilon f_{\theta}.
\end{equation}
In turn, making use of equation \eqref{eq:ch7_split_ecc_3_1_1} and \eqref{eq:ch7_split_ecc_3_2}, we find that the splitting probability starting from the outer disk is given by

\begin{equation}\label{eq:ch7_split_ecc_3}
	\pi_{\underline{R_1},R_2}(\theta)\underset{R_2,  R_1\gg  a_2	}{\sim}\pi^{(c)}_{\underline{R_1},R_2}(z_0-a_2 e^{i\theta}).
\end{equation}
Finally, making use of the continuous result $\pi^{(c)}$, we find that the asymptotic  splitting probability for a jump process starting from the outer rim at angle $\theta$ is exactly given by

\begin{equation}\label{eq:ch7_split_ecc_5}
	\pi_{\underline{R_1},R_2}(\theta)\underset{R_2,R_1\gg a_2	}{\sim}\frac{\log(\tilde{R}_2)-\log\left(\left \lVert\frac{c_2 + R_2(1-a_2) e^{i\theta}}{1+A(c_2 + R_2(1-a_2) e^{i\theta})}\right\rVert\right)}{\log(\tilde{R}_2)-\log(\tilde{R}_1)},
\end{equation}
\rev{
where  $A=d/\sqrt{(R_2^2-R_1^2)^2-2d^2(R_1^2+R_2^2)+d^4}$ and
\begin{equation}\label{eq:ch7_split_ecc_5_explicit}
	\tilde{R}_2=\frac{\sqrt{1+4R_2^2A^2}-1}{2R_2A^2}, \hspace{10pt} \tilde{R}_1=\frac{\sqrt{1+4R_1^2A^2}-1}{2R_1A^2}, \hspace{10pt} c_2 = \frac{A \tilde{R}_2^2}{1 - A^2 \tilde{R}_2^2}
\end{equation}
Of note, the discrete nature of the jump process is fully apparent via the emergence of the $a_2$ projected lengthscale. We note that, in the jointly asymptotic regime $R_1,R_2\gg a_2$ considered throughout, the splitting probability $\pi_{\underline{R_1},R_2}(\theta)$ is not $O(1)$: it converges to the corresponding Brownian splitting probability evaluated at a distance $o(R_2)$ from the outer rim, and vanishes logarithmically as $R_2\to\infty$, as also visible from the vertical axis of Fig.~\ref{fig:ch7_geom_disk}.}
To illustrate our analytical formulas in the $\mu=2$ case, we focus on the paradigmatic Run and Tumble (RTP) jump process, defined as an isotropic jump of exponentially distributed magnitude $|\boldsymbol{\ell}|\sim\text{Exp}(1/\gamma)$ and uniformly distributed direction\rev{, for which $a_2=1/(\sqrt{2}\gamma)$ in $d=2$ (see Appendix~\ref{app:RTP} for the derivation and its generalization to arbitrary $d$)}. The RTP has proved relevant in a variety of physical contexts \cite{Berg:2004,Tailleur:2008}.  Our method provides explicit expressions that fully quantify the splitting probability  for this process, as shown in Fig. \ref{fig:ch7_geom_disk}.

\subsection{\rev{Distribution of escape angles}}

\begin{figure}[h!]
	\centering
	\includegraphics[width=0.6\textwidth]{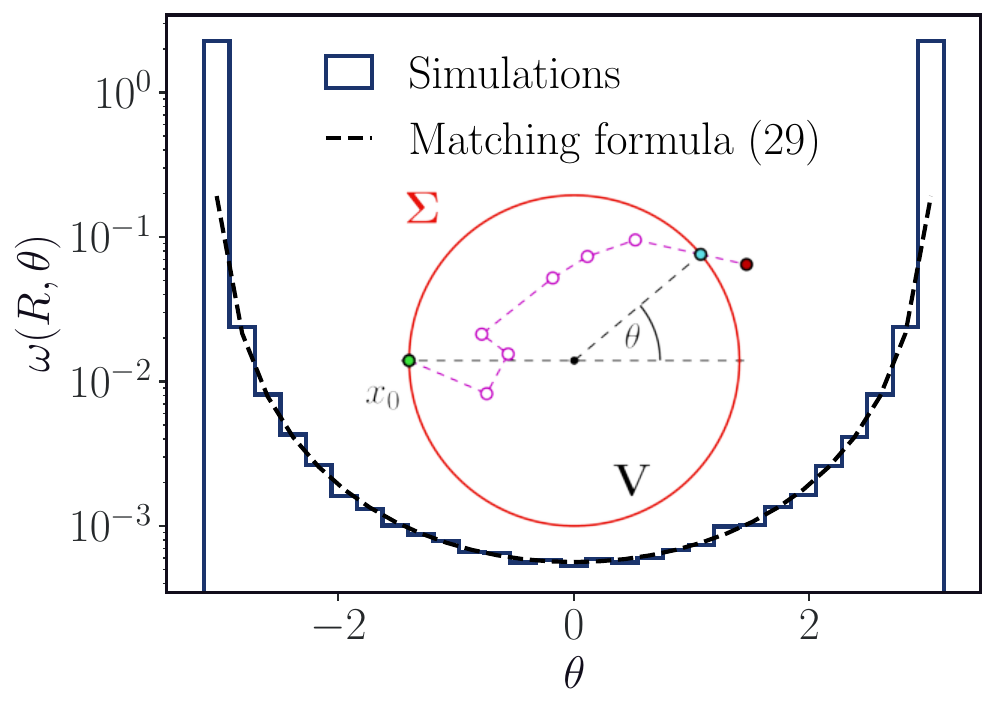}
	\caption{\rev{Distribution of escape angles of a RTP starting from the boundary of a disk ($r_0 = R,\theta_0=\pi$) with $\gamma=1$ and $R=100$. The dashed line is given by equation \eqref{eq:ch7_harmonic_measure_3} and the empirical histogram averaged over $10^6$ samples. 
	\textbf{(Inset)} This sample trajectory intersects the disk at an angle $\theta$ upon its first escape, contributing to $\omega(R,\theta)$.}}
	\label{fig:ch7_geom_disk_1}
\end{figure}

\rev{The matching formula is also applicable to spatially resolved FPO. In the large volume limit, the position of a particle with $\mu=2$ issued from $(r_0,\theta_0=\pi)$ upon crossing the absorbing disk surface is uniquely determined by the angular position $\theta$ (see Fig. \ref{fig:ch7_geom_disk_1}(a)), whose distribution $\omega(r_0,\theta)$ we refer to as the distribution of escape angles.}
Focusing on the specific case $r_0=R$, the asymptotic \rev{distribution of escape angles} is simply given by
 
 \begin{equation}\label{eq:ch7_harmonic_measure_1_1}
 	\omega(r_0,\theta)\underset{a_2 \ll R}{\sim}\pi^{\perp}_{0,\underline{a}}(R-r_0)\underset{u\to 0}{\lim} \left[\frac{\omega^{(c)}(R-u,R,\theta)}{\pi^{(c)}_{0,\underline{a}}(u)}\right],
 \end{equation}
\rev{where its continuous analog  $\omega^{(c)}(r,R,\theta)$, classically known as the Brownian harmonic measure, has the following expression \cite{Redner:2001a}}
\begin{equation}\label{eq:ch7_harmonic_measure_1}
	\omega^{(c)}(r_0,R,\theta)= \frac{1}{2\pi}\left[\frac{1-\frac{r_0^2}{R^2}}{1+\frac{2r_0}{R}\cos(\theta)+\frac{\rev{r_0^2}}{R^2}}\right].
\end{equation}

\noindent The $r_0\to R$ limit of $\omega^{(c)}(r_0,R,\theta)$ is straightforwardly extracted from equation \eqref{eq:ch7_harmonic_measure_1}, and we finally obtain the asymptotic \rev{distribution of escape angles} of the jump process starting from the boundary of the domain:

\begin{equation}\label{eq:ch7_harmonic_measure_3}
	\omega(R,\theta)\underset{R \to\infty}{\sim}\frac{1}{2\pi}\frac{a_2}{R}\left[\frac{1}{1+\cos(\theta)}\right].
\end{equation}
\rev{The comparison with numerical simulations of RTP is shown on Fig. \ref{fig:ch7_geom_disk_1}. 
While empirical histograms and the matching formula \eqref{eq:ch7_harmonic_measure_3} agree over a wide range of $\theta$, the empirical \rev{divergence} of histogram population around $\theta = \theta_0 = \pi$ should not come as a surprise. Indeed, in the large volume limit, most trajectories contributing to the empirical count close to the starting point do not penetrate the bulk of the domain.
In turn, the second term in equation \eqref{eq:ch7_stitching_3} cannot be neglected anymore and the matching formula does not hold in this case.
}

\section{$d$-dimensional Dynamical first-passage observables}\label{seq:5}

We now make use of the matching formula \eqref{eq:ch7_main_result} to compute the asymptotic behavior of paradigmatic dynamical observables, which have proven relevant in a number of physical contexts, from reaction kinetics \cite{VanKampen:1992} to foraging behavior \cite{Edwards2007}. Specifically, we first derive \rev{the mean exit time from a disk}, valid for both \rev{light-tailed (finite-variance, $\mu=2$)} and heavy-tailed processes, which so far have remained unsolved problems. Importantly, this prototypical example covers a broad range of physical situations, as it provides the mean residence time of \rev{a} particle in a confined domain.  We show that for processes with $\mu=2$ our procedure yields the complete distribution of the  first exit time. \rev{As in the previous sections, the confining domain boundary is here taken to be fully absorbing; Section~\ref{seq:6} extends this dynamical observable to confining domains with a reflecting boundary and an interior absorbing target.}

\subsection{\rev{Exit time from a disk}}

Following the previous discussion, we consider the mean exit time  $t(R)$ from a disk of radius $R$ for a particle issued from the boundary of the disk. To apply the matching formula, we make use of the continuous mean disk exit time for planar $\alpha$-stable processes \cite{Getoor1961}

\begin{equation}\label{eq:ch7_met_levy_1}
	t^{(c)}(R|r_0)=\left[2^\mu \Gamma\left(1+\frac{\mu}{2}\right)\Gamma\left(\frac{2+\mu}{2}\right)\right]^{-1}\left(R^2-r_0^2\right)^{\frac{\mu}{2}}.
\end{equation}
\rev{Replacing $\mathbb{P}(\ob(\xnh) = b)$ by $t(R|r_0)$ in the asymptotic equivalence \eqref{eq:ch7_main_result} we obtain the mean exit time for jump processes} 

\begin{equation}\label{eq:ch7_met_levy_2}
t(R)\underset{R \gg a_\mu }{\sim}\pi^\perp_{0,\underline{a}}(0)\underset{u \to 0}{\lim}\left[	\frac{t^{(c)}(R|R-u)}{\pi^{(c)}_{0,\underline{a}}(u)}\right].
\end{equation}

\begin{figure}[h!]
	\centering
	\includegraphics[width=0.6\textwidth]{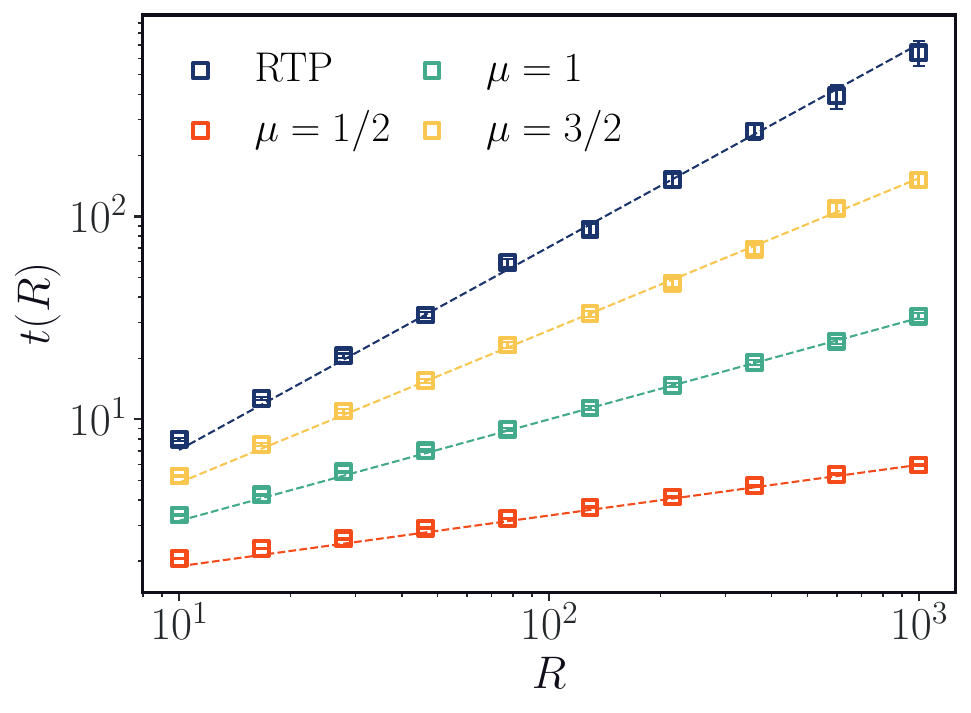}
	\caption{Mean exit-time from a disk of radius $R = 100$, starting from the boundary, for a RTP
	with $\gamma = 1$ and three planar heavy tailed jump processes with varying \rev{L\'evy} exponents and scale $1$. 
	\rev{
	The dotted lines correspond to the large $R$ matching method asymptotic result \eqref{eq:ch7_met_levy_2_bis_1}, using the explicit projected length scale $a_\mu$ derived in Appendix~\ref{app:RTP} and Appendix~\ref{app:radial_levy}
	}. The empirical estimates are averaged over $10^4$ particles evolved until the rim is crossed.}
	\label{fig:ch7_met_levy}
\end{figure}

\noindent In contrast to geometrical FPO, extra care needs to be taken in the identification of the limit process.
\rev{Indeed, the continuous mean first exit time \eqref{eq:ch7_met_levy_1} holds for $\alpha$-stable processes $\{\mathbf{x}_t\}_{t \in \mathbb{R}_+}$ defined via their Laplace transform $\mathbb{E}\left[e^{i \mathbf{k}\cdot \mathbf{x}_t} \right] = e^{-t |\mathbf{k}|^{\mu}}$.
To make use of the mean exit time matching formula \eqref{eq:ch7_met_levy_2} one needs to properly rescale the clock of the \rev{discrete} jump process, as detailed in Appendix \ref{app:projected}.
We illustrate the procedure in the case $\mu = 2$. The limit process is Brownian with diffusion coefficient $D=a_2^2$ and Laplace transform $\mathbb{E}\left[e^{i \mathbf{k}\cdot \mathbf{x}_t} \right] = e^{- D t |\mathbf{k}|^2}$. Using the scaling invariance of Brownian motion, the effective mean first exit time is given by $t^{(c)}_D \equiv t^{(c)} / D$. As a result, we obtain the asymptotic mean exit time
\begin{equation}\label{eq:ch7_met_levy_2_bis}
	t(R)  \underset{R \gg a_2 }{\sim}\frac{1}{D}\frac{a_2}{a}\underset{u \to 0}{\lim}\left[	\frac{ u R }{2 }\frac{ a}{u}\right] \underset{R \gg a_2 }{\sim} \frac{1}{a_2^2}\frac{a_2 R}{2} \underset{R \gg a_2 }{\sim} \frac{1}{2}\frac{R}{a_2}.
\end{equation}
For generic $\mu \in ]0,2]$, the time rescaling proceeds similarly with $t^{(c)}_{a_\mu} \equiv t^{(c)} / (a_\mu)^\mu$ yielding
\begin{equation}
	\label{eq:ch7_met_levy_2_bis_1}
	t^\mu(R) \underset{R \gg a_\mu}{\sim} \frac{1}{2^{\mu/2} \Gamma\left(1 + \frac{\mu}{2}\right)}\left[\frac{R}{a_\mu}\right]^{\frac{\mu}{2}}.
\end{equation}
Since the lengthscale $a_\mu$ depends on the complete jump distribution $p(\boldsymbol{\ell})$, we illustrate equation \eqref{eq:ch7_met_levy_2_bis_1} by computing in Appendix~\ref{app:RTP} and Appendix~\ref{app:radial_levy} the explicit $a_\mu$ for a RTP and a specific planar jump process whose increment $\boldsymbol{\ell} = \eta \boldsymbol{u}$ is such that the Fourier transform of the distribution $p(\eta)$ of $\eta$ is $\mathcal{F}\left[p(\eta)\right](k) = e^{-|k|^\mu}$ and $\boldsymbol{u}$ is drawn uniformly on the sphere. The agreement with numerical simulations is shown on Fig. \ref{fig:ch7_met_levy}.}

In the $\mu=2$ case, the matching procedure \rev{actually} yields more than the mean exit time, and we derive the complete asymptotic distribution of the exit time through the fully absorbing disk of radius $R$ for particles initiated from the rim. 
We emphasize that such results are crucial to quantify first passage time behavior in the context of surface mediated dynamics \cite{Calandre:2014ul}. Indeed, jump processes microscopically realize surface desorption events without the need to introduce some type of \textit{ad-hoc} ejection distance from the boundary.  

\begin{figure}[h!]
	\centering
	\includegraphics[width=0.6\textwidth]{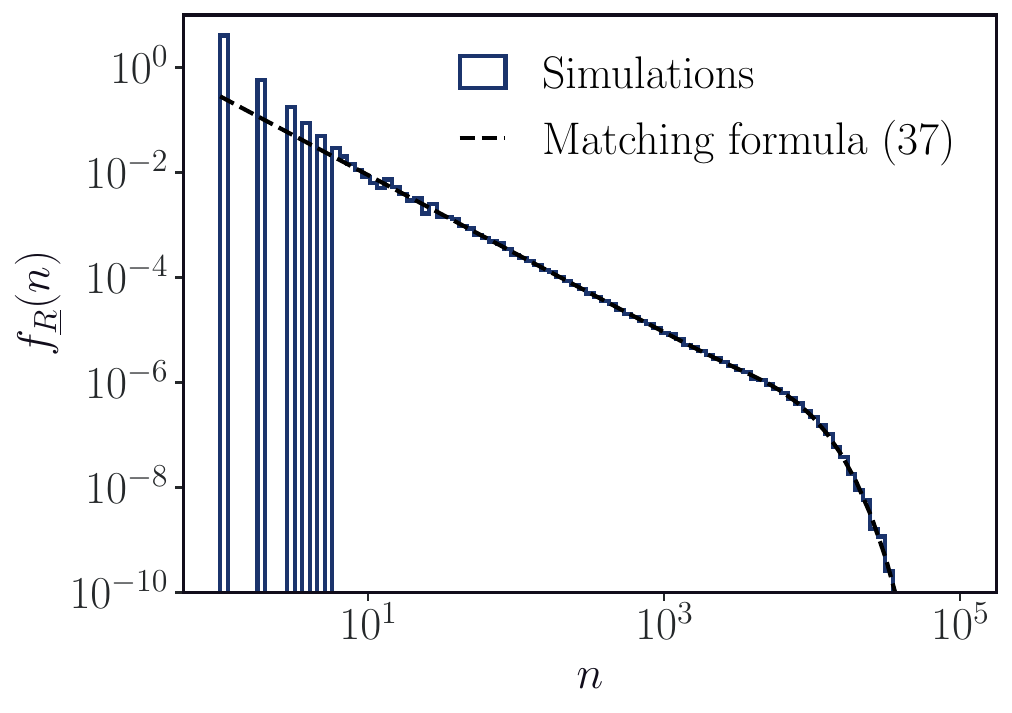}
	\caption{
	\rev{
	Distribution of the FPT $n$ over the boundary of a disk of radius $R=100$, for a RTP with $\gamma = 1$. The particle starts from the absorbing boundary, the theoretical dashed line is obtained from numerical inversion of equation \eqref{eq:ch7_fpt_to_rim_1_2} and the empirical histograms are averaged over $10^6$ particles evolved until the rim is crossed.  
	}
	}
\label{fig:ch7_complete_distrib}
\end{figure}

\rev{
We focus first on the first exit time distribution $f^{(c)}_{\underline{R}}(t|r)$ of a Brownian particle with diffusion coefficient $D$ starting from radius $r$ in an absorbing disk of radius $R$. This distribution satisfies the following diffusion equation
\begin{equation}\label{eq:ch7_fpt_to_rim_0_1}
	\frac{\partial}{\partial t}f^{(c)}_{\underline{R}}(t,r)=D\Delta_r\frac{\partial}{\partial t}f^{(c)}_{\underline{R}}(t,r)
\end{equation}
with boundary conditions $f^{(c)}_{\underline{R}}(t,r=R)=\delta(t)$, and can be explicitly computed in the Laplace transformed space:
\begin{equation}\label{eq:ch7_fpt_to_rim_1}
	\int_{0}^{\infty} e^{-s t}f^{(c)}_{\underline{R}}(t,r)\der t=\tilde{f}^{(c)}_{\underline{R}}(s,r)=\frac{I_0\left(\sqrt{\frac{s r^2}{D}}\right)}{I_0\left(\sqrt{\frac{s R^2}{D}}\right)}.
\end{equation}
To apply the matching formula, we seek to analyze the small $\epsilon$ limit of the edge initial condition $r = R-\epsilon$. Performing the expansion of the Laplace transform yields
\begin{equation}\label{eq:ch7_fpt_to_rim_1_1}
	\int_{0}^{\infty} e^{-s t}f^{(c)}_{\underline{R}}(t,R-\epsilon)\der t=\tilde{f}^{(c)}_{\underline{R}}(s,R-\epsilon)\underset{\substack{\epsilon\to 0}}{=}1-\epsilon\sqrt{\frac{s}{D}}\frac{I_1\left(\sqrt{\frac{R^2 s}{D}}\right)}{I_0\left(\sqrt{\frac{R^2 s}{D}}\right)}+o(\epsilon).
\end{equation}
The constant term in the right-hand side of equation \eqref{eq:ch7_fpt_to_rim_1_1} corresponds to the Dirac delta boundary condition,  such that we only consider the linearly vanishing term proportional to $\epsilon$. 
Building on the previous example of the mean first \rev{exit} time from a disk, the matching method amounts to replacing $\epsilon\to a_2$ and $D \to a_2^2$, yielding the discrete asymptotic first exit time distribution :
\begin{equation}\label{eq:ch7_fpt_to_rim_1_2}
	f_{\underline{R}}(n) \underset{R \gg a_2}{\sim} -{\rm InvLaplace}\left[\sqrt{s}\frac{I_1\left(\sqrt{\frac{R^2 s}{a_2^2}}\right)}{I_0\left(\sqrt{\frac{R^2 s}{a_2^2}}\right)}\right](n).
\end{equation}
Agreement with numerical simulations for a RTP particle is shown on Fig. \ref{fig:ch7_complete_distrib}. 
As in the example of escape angles, notice that the matching method fails to capture the \rev{distribution} at small values of $n$. This is expected, since events contributing to these empirical counts do not penetrate the bulk of the absorbing domain.
However, the approximation is good enough to capture the two typical behaviors of the first passage time distribution. The initial power law decay, with exponent $f_{\underline{R}}(n) \sim n^{-3/2}$ is \rev{reminiscent} of the particle being oblivious to the entire absorbing domain and locally interacting as if in a semi-infinite geometry, while the later time exponential decay is \rev{characteristic} of the finite volume survival probability suppression.
}

\section{Extension to reflecting boundary conditions ($\mu=2$)}\label{seq:6}

\begin{figure}[h!]
	\centering
	\begin{overpic}[height=0.4\textwidth]{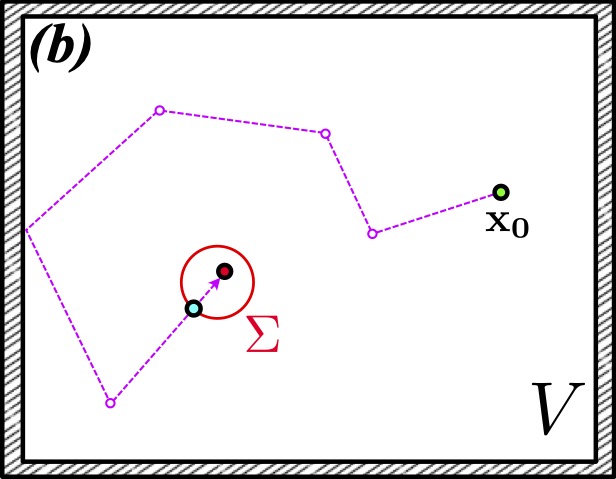}
		\put(5,65){\color{white}\rule{12\unitlength}{9\unitlength}}
	\end{overpic}
	\caption{
	\rev{
	The particle evolves in a domain with mixed boundary conditions. The outer boundary is \rev{reflecting}, \rev{while} the inner inclusion has absorbing boundary $\Sigma$. Jumps overshooting the reflecting boundary are reflected inside the domain in a billiard fashion. The particle evolves until it strictly crosses the boundary $\Sigma$ of the inner target domain. As in the eccentric splitting probability scenario, we neglect target overshoot events \rev{occurring} whenever a single jump leads the particle's \rev{trajectory to intersect $\Sigma$ twice}.
	This schematic is representative of the bulk initial condition, where the particle starts far from the absorbing target.
	}	
	}
	\label{fig:reflecting_schem}
\end{figure}

\rev{
Up to now, we restricted our attention to finite domains with a fully absorbing boundary. We now show that the framework can be extended to domains that additionally possess a reflecting portion of the boundary, as illustrated in Fig.~\ref{fig:reflecting_schem}. 
We restrict this extension to the case $\mu=2$. 
In this case, the reflecting boundary condition is unambiguously \rev{understood} both for the discrete jump process, and  at the level of the continuous-time limit process, defined as a reflected Brownian motion \cite{Skorokhod:1961,Whitt:2002}. 
For heavy-tailed jump processes with $\mu<2$, the meaning of a reflecting boundary condition is more subtle.
Indeed, incoming jumps over the reflecting boundary may have non-vanishing length relative to the system size, even in the large volume limit.
Understanding convergence to a limit $\alpha$-stable process also proves difficult as the mere reflecting boundary conditions for such processes are mathematically delicate objects, with several recent inequivalent constructions proposed in the literature \cite{Chechkin:2003,Garbaczewski:2019}.
Such convergence results are beyond the scope of this paper and we emphasize that the objective of the following discussion is to show that one can exploit existing large-volume results for reflected Brownian motion, rather than formally derive them. 
}

\rev{
To illustrate this point, we focus on the asymptotic first-passage time distribution to an interior absorbing target within a confining domain with reflecting outer boundary (see Fig.~\ref{fig:reflecting_schem}). 
}
This classical situation belongs to the class of target search problems, which has generated an intense activity  \cite{Redner:2001a,bookSid2014,Condamin2007,Benichou:2011fk,Benichou:2014fk,Levernier:2018qf,grebenkov2024target}. Specifically, the FPT distribution of a Brownian particle starting from $\xnh$ in a confining domain $\mathcal{D}$ of volume $V\propto R^d$ to a small spherical target of radius $r$ has been shown to display universal behavior in the large volume limit $R\gg r$ \cite{BenichouO.:2010,Meyer:2011}. More precisely, defining the mean first-passage time $t^{(c)}(\xnh)$ from $\xnh$ to the target, and the global mean first-passage time (GMFPT) $\langle \overline{T} \rangle = V^{-1}\int_\mathcal{D}t^{(c)}(\uh)\der \uh$, the distribution $f^{(c)}(\theta|\xnh)$ of the rescaled variable $\theta=t/\langle \overline{T} \rangle$  \rev{
is shown in \cite{BenichouO.:2010,Benichou:2014fk} to be asymptotically given by}

\begin{equation}\label{eq:ch7_fpt_last_1}
	f^{(c)}(\theta|\xnh)\underset{R\to\infty}{\sim}\left(1-\frac{t^{(c)}(\xnh)}{\langle \overline{T} \rangle}\right)\delta(\theta)+\frac{t^{(c)}(\xnh)}{\langle \overline{T} \rangle}e^{-\theta}.
\end{equation}
\rev{
We consider the case of a jump process issued from the surface of the interior target and denote $f(\theta)$ the corresponding rescaled FPT $\theta=n/\langle\overline{T}\rangle$ distribution. Being a continuous quantity, the continuous mean FPT $t^{(c)}(\xnh)$ vanishes as the source target distance goes to 0, such that equation \eqref{eq:ch7_fpt_last_1} cannot be used directly to evaluate $f(\theta)$.
As done previously, we make use of the matching formula  \eqref{eq:ch7_main_result} to obtain the asymptotic equivalence 
}

\begin{equation}\label{eq:ch7_asymptotic_fpt}
	f(\theta)\underset{\substack{R\gg a_2 \\ n\gg1}}{\sim}\pi^\perp_{0,\underline{a}}(0)\underset{\xnh \to \Sigma}{\lim}\left[\frac{f^{(c)}(\theta|\xnh)}{\pi^{(c)}_{0,\underline{a}}(\xnh)}\right],
\end{equation} 
where $\Sigma$ denotes the surface of the absorbing target. 
Consequently, the evaluation of  $t^{(c)}$ and $\langle \overline{T} \rangle$ as $\xnh\to \Sigma$ is sufficient to obtain the complete FPT distribution $f(\theta)$. 
As an illustration, we consider a three-dimensional domain $\mathcal{D}$ of arbitrary shape, for which the large volume behavior of $t^{(c)}(\xnh)$ and $\langle \overline{T} \rangle$ read \cite{Condamin:2007eu}

\begin{equation}\label{eq:ch7_fpt_last_2}
	t^{(c)}(\xnh)\underset{R\to\infty}{\sim}\frac{V}{4\pi D}\left(\frac{1}{r}-\frac{1}{r+\epsilon}\right),\hspace{20pt}\langle \overline{T} \rangle\underset{R\to\infty}{\sim}\frac{V}{D}\frac{1}{4\pi r},
\end{equation}

\noindent with $r$ the target radius, and $\epsilon$ the distance of the Brownian particle to the target surface. Importantly, these expressions only depend on the volume $V$, and not on the specific shape of the confining volume. Since  $t^{(c)}(r+\epsilon)$ vanishes linearly with $\epsilon$ we combine equation \eqref{eq:ch7_asymptotic_fpt} and \eqref{eq:ch7_fpt_last_2} to obtain the distribution of the rescaled variable $\theta$ in the large $R$ limit

\begin{equation}\label{eq:ch7_fpt_last_4_1}
	f(\theta)\underset{R\to\infty}{\sim}\left(1-\frac{a_2}{r}\right)\delta(\theta)+\frac{a_2}{r}e^{-\theta}.
\end{equation}
\begin{figure}[h!]
\centering
	\includegraphics[width=1.0\textwidth]{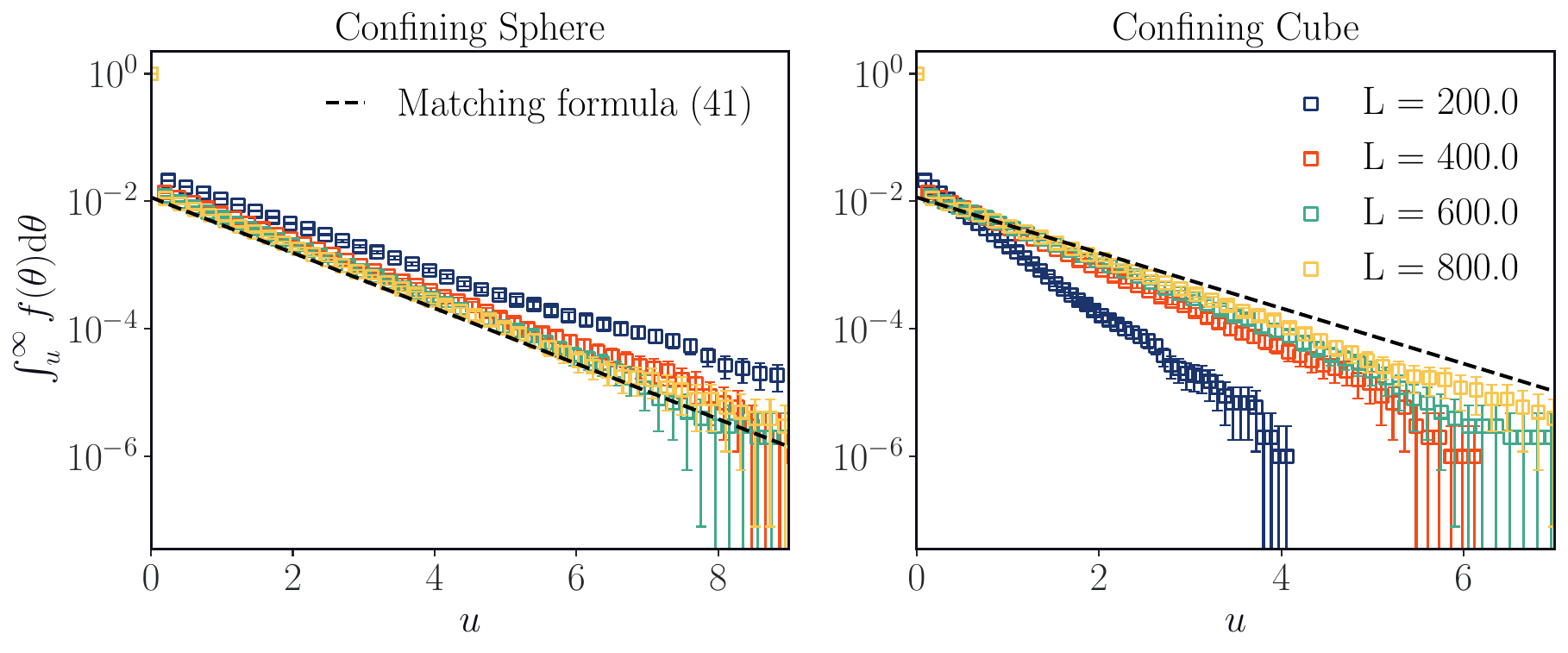}
\caption{
\rev{
Cumulative distribution of the rescaled FPT $\theta=n/\langle \overline{T}\rangle$ to a centered spherical target of radius $r=50$ in a three-dimensional confining sphere (left-hand side) or cube of edge length $L$ (right-hand side). 
The spherical domain's confining radius is chosen such that its volume matches the corresponding confining cube.
The particle starts on the boundary of the inner target and follows Run and Tumble motion with $\gamma = 1$.
In the confining sphere scenario, the exact analytical GMFPT $\langle \overline{T}\rangle$ is used to compute rescaled first passage times, while in the confining cube setup we use the large volume asymptotic formula \eqref{eq:ch7_fpt_last_2}.
The dashed theoretical line is given by the cumulative distribution of equation \eqref{eq:ch7_fpt_last_4_1}.
Empirical counts are averaged over $10^6$ particles evolved until the inner target is reached.
}
}
\label{fig:ch7_complete_distrib_2}
\end{figure}

\noindent \rev{
It appears clearly that the rescaled FPT distribution inherits the asymptotic functional form of its continuous counterpart, while the prefactors directly involve the discrete lengthscale $a_2$.
The theoretical prediction \eqref{eq:ch7_fpt_last_4_1} is tested for both spherical and cubic confining geometries as shown in Fig. \ref{fig:ch7_complete_distrib_2}.}
\rev{For spherical confinement we use the exact GMFPT to compute rescaled first passage time and observe perfect agreement.
In the cubic case, no exact GMFPT \rev{expression} is \rev{available} and we make use of the asymptotic approximation \eqref{eq:ch7_fpt_last_2}.
\rev{Convergence} of the cumulative distribution of the rescaled FPT towards the \rev{exponential} is clear but slow. Note however that the objective of the matching method is to capture the prefactor which is achieved - the mismatch of the exponential decay rate is rather a consequence of the large volume asymptotic behavior of the \rev{GMFPT}.
In short, the matching procedure is fully compatible with existing large volume frameworks to compute continuous FPO statistics, and allows us to extend numerous asymptotic results to jump processes, while retaining the specificity arising from their \rev{discreteness}.
}

\section{Conclusion}

Jump processes offer an alternative description of single particle stochastic dynamics that encapsulate inherent discrete effects naturally arising - for example - in experimental sampling procedures. In turn, understanding the impact of such discrete nature on stochastic observables is key to correctly interpret empirical time series. However, because of this very \rev{discreteness}, analytical results of jump process observables are sparse.

In this paper, we focused on general first passage observables evaluated upon first exit from a confining domain, which have proved to be relevant in a variety of experimental contexts, from biology \cite{Blanco:2003a} to photonic transport \cite{Savo:2017, Baudouin:2014}. By leveraging the convergence of jump processes towards well-identified continuous limit processes, we developed a matching framework to systematically evaluate large volume behavior of FPO statistics in all initial position regimes, provided that corresponding continuous expressions be known. Importantly, we have shown that the edge
regime, where particles start close to the absorbing boundary, is of particular relevance, and our framework safeguards the discrete signature of jump processes, even in the large confining volume limit.

In that context, we derived a variety of novel expressions for various geometrical and dynamical FPO of jump processes\rev{, restricting our attention, in Sections \ref{seq:3}--\ref{seq:5}, to confining domains with a fully absorbing boundary}. In particular, we first showed that our method is valid in non symmetrical domains by computing eccentric splitting probabilities in disk geometries. We then showcased its broad application range by computing mean exit times for heavy-tailed processes in disks\rev{. Finally, restricting to processes with $\mu=2$, for which reflecting boundary conditions are unambiguously defined, we showed in Section \ref{seq:6} that the matching procedure can be extended to existing continuous large volume first passage frameworks for confining domains with a reflecting boundary, by computing asymptotic FPT distributions to absorbing interior targets; extending this last result to heavy-tailed ($\mu<2$) processes, for which reflecting boundary conditions are considerably more subtle, is left for future work}.

 \begin{acknowledgments}
 R.V. acknowledges support of ERC synergy grant SHAPINCELLFATE. 
\end{acknowledgments} 

\appendix

\section{On isotropic jump processes}\label{app:projected}

\subsection{Limit process}

In this section, we give some details on the different definitions of isotropic jump processes used throughout the text. Note that this is not a formal account, rather a reminder of the different formulas we have used in the calculations. \\

For a d-dimensional jump process $\{\mathbf{x}_n\}_{n \in \mathbb{N}} \in \mathbb{R}^d$, we denote the single jump distribution $p(\boldsymbol{\ell})$. The Fourier transform of the single jump distribution is given by

\begin{equation}
	\tilde{p}(\boldsymbol{k}) = \int_\mathbb{R}^d \der \ell p(\boldsymbol{\ell}) e^{i \boldsymbol{k}\cdot \boldsymbol{\ell}}.
\end{equation}
We consider a class of isotropic jump processes, whose $\tilde{p}(\boldsymbol{k})$ small $|\boldsymbol{k}|$ behavior reflects the power law behavior of the jump length distribution 

\begin{equation}
	\tilde{p}(\boldsymbol{k}) \underset{|\boldsymbol{k}| \to 0}{=} 1 - (a_\mu|\boldsymbol{k}|)^\mu + o(|\boldsymbol{k}|).
\end{equation}
The isotropy is reflected in the \rev{dependence} of $\tilde{p}$ on the norm of $\boldsymbol{k}$ only. The {L\'evy} exponent $\mu$ \rev{characterizes} the algebraic decay of the tails of the jump length distribution, and together with the jump length scale $a_\mu$, define the limit continuous process. Informally, the large $n$ Fourier transform of the distribution of the particle after $n$ steps is given by

\begin{equation}
	\tilde{p}(\boldsymbol{k},n) \simeq e^{- n (a_\mu |\boldsymbol{k}|)^\mu}
\end{equation}
Identifying steps with continuous time, one obtains the limit continuous process :

\begin{itemize}
	\item When $\mu=2$ the jump process converges towards the $d$-dimensional Brownian motion with $D=a_2^2$ and instantaneous characteristic function
	\begin{equation}
		\mathbb{E}\left[e^{i \boldsymbol{k}\cdot \boldsymbol{x}_t}\right] = e^{-t D |\boldsymbol{k}|^2}
	\end{equation}
	\item When $\mu\in ]0,2[$ the jump process converges towards the $d$-dimensional scaled $\alpha$-stable process with instantaneous characteristic function
	\begin{equation}
		\mathbb{E}\left[e^{i \boldsymbol{k}\cdot \boldsymbol{x}_t}\right] = e^{-t (a_\mu |\boldsymbol{k}|)^\mu}
	\end{equation}

\end{itemize}

\subsection{Projected Jump process}

The projected jump process is the effective one-dimensional process obtained by projecting the isotropic increments $\boldsymbol{\ell}$ onto a single fixed direction, say $\boldsymbol{e}_1$. The corresponding jump process $\{x^\perp_n\}$ has the one-dimensional single jump distribution

\begin{equation}
	p_\perp(\ell) = \int_{\mathbb{R}^d} p(\boldsymbol{\ell})\delta(\boldsymbol{\ell}\cdot \boldsymbol{e}_1 - \ell)\der \boldsymbol{\ell},
\end{equation}
\rev{with $\ell \in \mathbb{R}$}. Note that by virtue of the isotropy of the original process, the projected jump process is symmetric, and inherits the tail behavior and L{\'evy} exponent. As such, we define the projected length scale $a_\mu^\perp$ via the small $|k|$ behavior of the Fourier transform of $p_\perp(\ell)$

\begin{equation}
	\tilde{p}_\perp(k) = 1 - (a_\mu^\perp |k|)^\mu + o(|k|^\mu)
\end{equation}
Let us further show that $a_\mu^\perp = a_\mu$. Starting from projected distribution one has

\begin{equation}
	\begin{split}
		\int_\mathbb{R} e^{i k \ell}p_\perp(\ell)\der \ell &= \int_\mathbb{R} e^{i k \ell} \left[\int_{\mathbb{R}^d}p(\boldsymbol{\ell})\delta(\boldsymbol{\ell}\cdot \boldsymbol{e}_1 - \ell)\der \boldsymbol{\ell}\right]\der \ell\\
		&= \int_{\mathbb{R}^d} e^{i k \boldsymbol{\ell}\cdot \boldsymbol{e}_1
		}p(\boldsymbol{\ell})\der \boldsymbol{\ell}
	\end{split}
\end{equation}
such that
\begin{equation}
	\tilde{p}_\perp(k) = \tilde{p}(k \boldsymbol{e}_1) \implies (a_\mu^\perp - a_\mu) |k|^\mu \underset{|k|\to 0}{=} o(|k|^\mu) \implies a_\mu = a_\mu^\perp.
\end{equation}
In turn, computing $a_\mu^\perp$ for isotropic processes is sufficient to identify $a_\mu$ and the corresponding limit process. 

\subsection{Alternative parametrization}

We end this section with an alternative parametrization \rev{and sampling procedure} of isotropic distributions as it will be useful for the specific jump process examples we consider below. The increment $\boldsymbol{\ell}$ is sampled in two steps : 
\begin{itemize}
\item First, choose a direction $\boldsymbol{u}$ at random on the sphere.
\item Second, draw a one-dimensional real random variable \rev{$\eta \in \mathbb{R}$} from a symmetric distribution $p(\eta)$, which we will refer to as {\bf the signed jump length.}
\item The isotropic increment is then obtained as $\boldsymbol{\ell} = \eta \boldsymbol{u}$.
\end{itemize}
This representation is particularly practical in identifying the distribution $p_\perp$ of the \rev{projected jump $\ell \in \mathbb{R}$.} Indeed, the projected distribution in dimension $d$ can be rewritten as (see \cite{Mori2020a} for instance):

\begin{equation}
\label{eq:app_proj}
p_\perp(\ell)=\frac{\Gamma(d/2)}{\sqrt{\pi}\Gamma((d-1)/2)}\int_0^\infty\frac{1}{\eta}\left(1-\left(\frac{\ell}{\eta}\right)^2\right)^{(d-3)/2}\Theta\left(1-\left|\frac{\ell}{\eta}\right|\right)2 p(\eta) \text{d}\eta.
\end{equation}
In the following section, we use the relation \eqref{eq:app_proj} to derive \rev{$a_\mu$} for two distinct jump processes.

\section{Run and Tumble processes}\label{app:RTP}

\subsection{Projected distributions}

\rev{Recall from Section~\ref{seq:4} that the RTP is an isotropic jump process of exponentially distributed magnitude and uniform direction; equivalently, it can be written as $\boldsymbol{\ell}=\eta \boldsymbol{u}$ with $\boldsymbol{u}$ uniform on the unit sphere and $\eta$ the signed jump length,} distributed according to

\begin{equation}
	p(\eta) = \frac{\gamma}{2} e^{-\gamma |\eta|}
\end{equation}

\begin{itemize}
	\item In $d=1$, the projected jump distribution is simply \rev{given} by the signed jump length
	
	\begin{equation}
		p_\perp(\ell) = \frac{\gamma}{2} e^{-\gamma |\ell|}
	\end{equation}
	from which one obtains $a_2 = 1 / \gamma$ by computing the variance of $\ell$

	\item In $d=2$, the projected jump distribution is given by
	
	\begin{equation}
		p_\perp(\ell) =  \frac{\gamma}{\pi}K_0\left(\gamma|\ell|\right)
	\end{equation}
	where $K_0$ is the 0th order modified Bessel function of the second kind. In turn, one obtains $a_2 = 1 / (\sqrt{2}\gamma)$ by computing the variance of $\ell$.

	\item In $d=3$, the projected jump distribution is given by
	
	\begin{equation}
	p_\perp(\ell) = \frac{\gamma}{2}\Gamma\left(0,\gamma|\ell|\right)
	\end{equation}	
	 where $\Gamma(0,b)$ is the 0th order incomplete gamma function. In turn, one obtains $a_2 = 1 / (\sqrt{3}\gamma)$ by computing the variance of $\ell$.

\end{itemize}

\subsection{A generic note on \emph{Physical} Run and Tumble processes with speed $v_0$}

So far, we have considered Run and Tumble processes through the lens of exponentially distributed jump lengths, i.e. neglecting the physical ``speed'' of the particle and directly assigning an exponential distribution to each jump length.
More physically, RTP motion is often described as a particle moving ballistically at constant speed $v_0$ for a random run duration $\tau$, distributed according to $q(\tau)$ (not necessarily exponential), before tumbling to a new, independent, uniformly distributed direction. Provided $q(\tau)$ has finite second moment $\langle \tau^2 \rangle$, this construction is an instance of the alternative parametrization of Appendix~\ref{app:projected}: the signed jump length is $\eta = \pm v_0 \tau$ (sign chosen with probability $1/2$), so that $\mathbb{E}[\eta^2] = v_0^2 \langle \tau^2 \rangle$, combined with a direction $\boldsymbol{u}$ drawn uniformly on the sphere.

For such an isotropic increment $\boldsymbol{\ell} = \eta \boldsymbol{u}$, the small $|\boldsymbol{k}|$ expansion of the Fourier transform reads $\tilde{p}(\boldsymbol{k}) = \mathbb{E}\left[e^{i \boldsymbol{k}\cdot\boldsymbol{\ell}}\right] \underset{|\boldsymbol{k}|\to 0}{\sim} 1 - \frac{1}{2}\mathbb{E}\left[(\boldsymbol{k}\cdot\boldsymbol{\ell})^2\right]$, and since $\boldsymbol{u}$ is uniform on the $(d-1)$-sphere, $\mathbb{E}\left[(\boldsymbol{k}\cdot\boldsymbol{u})^2\right] = |\boldsymbol{k}|^2/d$, so that

\begin{equation}
	\mathbb{E}\left[(\boldsymbol{k}\cdot\boldsymbol{\ell})^2\right] =  \frac{|\boldsymbol{k}|^2}{d}\,v_0^2\langle \tau^2 \rangle.
\end{equation}

\noindent Comparing with $\tilde{p}(\boldsymbol{k})\underset{|\boldsymbol{k}|\to 0}{=}1-(a_2|\boldsymbol{k}|)^2+o(|\boldsymbol{k}|^2)$ (Eq.~\eqref{eq:ch7_ddim_jp} at $\mu=2$), we obtain, for general dimension $d$,

\begin{equation}\label{eq:ch7_fpt_to_rim_4}
	a_2 = v_0\sqrt{\frac{\langle \tau^2 \rangle}{2d}}.
\end{equation}

\noindent Explicitly, for $d=1,2,3$:

\begin{equation}
	a_2 = v_0\sqrt{\frac{\langle \tau^2 \rangle}{2}}, \hspace{15pt} v_0\sqrt{\frac{\langle \tau^2 \rangle}{4}}, \hspace{15pt}v_0\sqrt{\frac{\langle \tau^2 \rangle}{6}}\hspace{15pt}(d=1,2,3\text{ respectively}).
\end{equation}

\noindent As a consistency check, specializing to exponential waiting times $q(\tau)=\gamma e^{-\gamma \tau}$ (so that $\langle \tau^2\rangle = 2/\gamma^2$) and $v_0=1$ recovers exactly $a_2 = 1/\gamma, 1/(\sqrt{2}\gamma), 1/(\sqrt{3}\gamma)$ for $d=1,2,3$, matching the values obtained directly above from the projected distributions $p_\perp(\ell)$.

In this physical setting, one needs to be cautious when evaluating the effective diffusion coefficient of the limit \rev{Brownian} motion. Indeed, in the identification $D=a_2^2$ defined in equation \eqref{eq:ch7_fpt_to_rim_4}, $D$ is the diffusion coefficient conjugate to the discrete \emph{step count} $n$ and \rev{implicitly} assumes that each intertumbling waiting time is equal to one unit of the physical time.
For a physical Run and Tumble particle, however, each step consumes a random physical duration $\tau_i$, with mean $\langle \tau \rangle$, generally different from $1$. The  time elapsed after $n\gg1$ tumbling events is $t\approx n\langle \tau \rangle$, so that the true physical diffusion coefficient $D_{\rm phys}$, defined through $\langle |\boldsymbol{x}(t)|^2\rangle \underset{t\to\infty}{\sim} 2dD_{\rm phys}\,t$, with $t$ real (clock) time, is related to $v_0$ and $\tau$ by

\begin{equation}\label{eq:rtp_dphys}
	D_{\rm phys} = \frac{a_2^2}{\langle \tau \rangle} = \frac{v_0^2\langle \tau^2\rangle}{2d\langle \tau \rangle}.
\end{equation}
Equivalently, any mean first-passage \emph{step count} $t(R)$ obtained via the matching method translates into a physical first-passage \emph{time} as $t_{\rm phys}(R) \approx \langle \tau \rangle\, t(R)$.

\section{Radial L\'evy processes}\label{app:radial_levy}

\rev{Recall from Section~\ref{seq:5} the radial Levy process introduced via the Fourier transform of their signed jump length distribution $\tilde{p}(k) = e^{-(c |k|)^\mu}$.} We now extract the relevant projected length scale $a_\mu$ in $d=2,3$.

\subsection{Two dimensions}
	
In two dimensions, we compute directly the Fourier transform of the projected jump process. Introducing the 0th order Bessel function of the first kind $J_0$, we have:

\begin{equation}
	\begin{split}
		\Tilde{p}_\perp(k)&=\frac{1}{\pi}\int_{-\infty}^\infty e^{ik \ell}\left[\int_0^\infty\frac{1}{\eta}\left(1-\left(\frac{\ell}{\eta}\right)^2\right)^{-\frac{1}{2}}\theta\left(1-\left|\frac{\ell}{\eta}\right|\right)2p(\eta)\text{d}\eta\right]\text{d}\ell\\
		&=\int_0^\infty J_0(k\eta)2p(\eta)\text{d}\eta \\
		&=\int_{-\infty}^\infty J_0(k\eta)p(\eta)\text{d}\eta \\
		&=\int_{-\infty}^\infty \frac{1}{\pi}\int_0^\pi e^{i k \eta \cos(\theta)}p(\eta)\text{d}\theta \text{d}\eta \\
		&=\frac{1}{\pi}\int_{0}^\pi \Tilde{p}(k\cos(\theta))\text{d}\theta.
	\end{split}
\end{equation}
In turn, we obtain the small $k$ expansion of $\tilde{p}_\perp(k)$:
\begin{equation}
	\Tilde{p}_\perp(k)\underset{k\to 0}{=}1-\frac{2}{\pi}|c k|^\mu\int_0^\frac{\pi}{2}\cos(\theta)^\mu\text{d}\theta +o(k^\mu) 
\end{equation}
yielding the following value for $a_{\mu}$ :
\begin{equation}
	a_\mu=c\left(\frac{\Gamma(\frac{1+\mu}{2})}{\sqrt{\pi}\Gamma(1+\frac{\mu}{2})}\right)^{\frac{1}{\mu}}
\end{equation}

\subsection{Three dimensions}

Similarly in three dimensions, we compute directly the Fourier transform of $p_\perp(\ell)$:

\begin{equation}\label{eq:ch4_proj_1}
	\begin{split}
		\Tilde{p}_\perp(k)&=\frac{1}{2}\int_{-\infty}^\infty e^{ik \ell}\left[\int_0^\infty\frac{1}{\eta}\theta\left(1-\left|\frac{\ell}{\eta}\right|\right)2p(\eta)\text{d}\eta\right]\text{d}\ell\\
		&=\int_{0}^\infty \frac{\sin(k\eta)}{k\eta}2p(\eta)\text{d}\eta\\
		&=\frac{1}{k}\int_{-\infty}^\infty \frac{\sin(k \eta)}{\eta}p(\eta)\text{d}\eta\\
	\end{split}
\end{equation}

\noindent We now use the specific form of the 3D jump distribution $\tilde{p}(k)=e^{-|c k|^\mu}$ and write $\sin(k\eta)=\text{Im}(e^{ik\eta})$:

\begin{equation}\label{eq:ch4_proj_2}
	\begin{alignedat}{1}
		\Tilde{p}_\perp(k)&=\frac{1}{k}\int_{-\infty}^\infty \frac{e^{ik\eta}}{\eta}\left[\frac{1}{2\pi}\int_{-\infty}^\infty e^{-ik'\eta-|c k'|^\mu}\text{d}k'\right]\text{d}\eta \\
		&=\frac{1}{k}\int_{-\infty}^\infty \int_{-\infty}^\infty \frac{e^{i\eta(k-k')}}{\eta}\text{d}\eta\frac{1}{2\pi} e^{-|c k'|^\mu}   \text{d}k' \\
		&=\frac{i}{k}\int_{-\infty}^\infty \frac{1}{2}\text{sign}(k-k')e^{-| c k'|^\mu}\text{d}k'. \\
	\end{alignedat}
\end{equation}

\noindent Taking the imaginary part finally yields:

\begin{equation}\label{eq:ch4_proj_3}
	\Tilde{p}_\perp(k)=\frac{\Gamma\left(\frac{1}{\mu}\right)-\Gamma\left(\frac{1}{\mu},(c \rev{|k|})^\mu\right)}{c \mu \rev{|k|}}
\end{equation}

\noindent with $\Gamma(s,x)$ the incomplete gamma function. The small $|k|$ asymptotic analysis leads to

\begin{equation}\label{eq:ch4_proj_4}
	\begin{alignedat}{1}
		&	\Tilde{p}_\perp(k)\underset{k\rightarrow0}{\sim}1-\frac{(c \rev{|k|})^\mu}{1+\mu}+o(k^\mu)\\
	\end{alignedat}
\end{equation}
and we find

\begin{equation}
	a_\mu=\frac{c}{(1+\mu)^{1/\mu}}.
\end{equation}

\section{On the eigenfunction decomposition of continuous first-passage observables}\label{app:eigen}

{
\color{black}
\noindent We focus here on the existence of the limit on the right-hand side of equation \eqref{eq:ch7_main_result}, which we recall:

\begin{equation}\label{eqSM:ch7_main_result}
	\left[\frac{\mathbb{P}(\ob(\xnh)=b)}{\pi^\perp_{0,\underline{a}}(d_e)}\right]
	\underset{d_e\lesssim a_\mu \ll a \ll R}{\sim}
	\left[\frac{\mathbb{P}(\ob^{(c)}(\xnh)=b)}{\pi^{(c)}_{0,\underline{a}}(d_e)}\right]
	\underset{\substack{R \to \infty \\ \mathbf{u} \to \mathbf{x_h}}}{\sim} \left[\frac{\mathbb{P}(\ob^{(c)}(\mathbf{u})=b)}{\pi^{(c)}_{0,\underline{a}}(\mathbf{u})}\right].
\end{equation}
The key assumption we make is that all the distributions of continuous first-passage observables can be projected on a basis $\{\psi_k\}$ of eigenfunctions of the generator $\mathcal{L}$ of the continuous limit process: the ordinary Laplacian $\mathcal{L}=\Delta$ for $\mu=2$, and the fractional Laplacian $\mathcal{L}=-(-\Delta)^{\mu/2}$ for $\mu\in]0,2[$, with Dirichlet boundary conditions on the absorbing surface $\Sigma$, \textit{i.e.}, $\mathcal{L}\psi_k=-\lambda_k\psi_k$ with $\psi_k|_{\Sigma}=0$:

\begin{equation}
	P(\ob^{(c)}(\mathbf{u})=b)= \sum_{k=1}^{\infty} c_k \psi_k(\bo{u}),
\end{equation}
\noindent where the $c_k$ depend on both $\ob^{(c)}$ and $b$. In turn, the leading $\bo{u}$ behavior of $P(\ob^{(c)}(\mathbf{u})=b)$ as $\bo{u}\to \bo{x_h}$ is identical for all first-passage observables, and in particular for the splitting probability $\pi^{(c)}_{0,\underline{a}}(\mathbf{u})$, such that the above-defined ratio is non-vanishing.
Importantly, we expect the eigenfunction decomposition to hold for observables satisfying inhomogeneous fractional Laplace equations, which we assume is the case for all observables discussed in the main text. 

We emphasize that this argument is heuristic, in that rigorous spectral results for the fractional Laplacian with Dirichlet boundary conditions 
are, to our knowledge, only firmly established for specific one-dimensional geometries such as the interval \cite{Kwasnicki:2012}, and not for the general (and generally higher-dimensional) confining domains considered in this paper. Nevertheless, we expect this qualitative picture 
to extend to the smooth bounded domains $\mathcal{D}$ considered here.
}

\section{On the eccentric splitting probability for Brownian motion in a disk}\label{app:eccentric}

In this section we focus on the splitting probability of a Brownian particle in the eccentric geometry depicted in Fig.~\ref{fig:ch7_geom_disk}. Precisely, we define $\pi^{(c)}_{R_1,\underline{R_2}}(z)$ the probability that the particle reaches the outer rim before the inner one, starting from a position $z$.   To compute $\pi^{(c)}_{R_1,\underline{R_2}}(z)$ explicitly, we first introduce the center distance $d$ and define the constant $A=d/\sqrt{(R_2^2-R_1^2)^2 -2d^2(R_1^2+R_2^2) +d^4}$
\rev{
and the effective radii
\begin{equation}\label{eq:chap7_split_ecc_4}
		\begin{alignedat}{1}
			&\tilde{R}_2=\frac{\sqrt{1+4R_2^2 A^2}-1}{2 R_2 A^2}\\
			&\tilde{R}_1=\frac{\sqrt{1+4R_1^2 A^2}-1}{2 R_1 A^2}.
		\end{alignedat}
	\end{equation}
and eccentric centers
\begin{equation}
		c_i= \frac{A \tilde{R}_i^2}{1 - A^2 \tilde{R}_i^2}.
\end{equation}
In turn, the conformal transformation
\begin{equation}\label{eq:chap7_split_ecc_1}
		w(z)=\frac{(z + c_2)}{1+A (z + c_2)}
\end{equation}
maps the eccentric region $\{z; |z-l| \geq R_1, |z|\leq R_2\}$  to a concentric region between two disks of radii $\tilde{R}_2>\tilde{R}_1$ (see \cite{Chen:2009} for more details). 
}

Next, since the Laplace equation satisfied by $\pi^{(c)}$ is invariant by conformal transformation, if $w(z)\in \mathbb{C}$ is the image of $z$ through the conformal mapping,  $\pi^{(c)}_{R_1,\underline{R_2}}(z)= \pi^{(c)}_{\tilde{R}_1,\underline{\tilde{R}_2}}(w(z))$. Finally, in the concentric geometry, the splitting probability only depends on the initial radius $r\in [\tilde{R}_1,\tilde{R}_2]$, and the splitting probability is known \cite{Redner:2001a}:
	
\begin{equation}\label{eq:ch7_split_ecc_4}
	\pi^{(c)}_{\tilde{R}_1,\underline{\tilde{R}_2}}(r)=\frac{\log\left(r\right)-\log(\tilde{R}_1)}{\log(\tilde{R}_2)-\log(\tilde{R}_1)}.
\end{equation}
such that	

\begin{equation}\label{eq:app_split_ecc_final}
	\pi^{(c)}_{R_1,\underline{R_2}}(z)=\frac{\log\left(|w(z)|\right)-\log(\tilde{R}_1)}{\log(\tilde{R}_2)-\log(\tilde{R}_1)},
\end{equation}
and 

\begin{equation}\label{eq:ch7_split_ecc_6}
	\pi^{(c)}_{\underline{R_1},R_2}(z)=\frac{\log(\tilde{R}_2)-\log\left(|w(z)|\right)}{\log(\tilde{R}_2)-\log(\tilde{R}_1)}.
\end{equation}
\rev{
We finally comment on the small $\epsilon$ asymptotic behavior of $\pi^{(c)}_{\underline{R_1}, R_2}((R_2 - \epsilon)e^{i\theta})$ as the initial particle position tends towards the outer absorbing rim. Formally, the small $\epsilon$ expansion reads
\begin{equation}
	\pi^{(c)}_{\underline{R_1}, R_2}((R_2 - \epsilon)e^{i\theta}) = \frac{\epsilon}{\log(\tilde{R}_2)-\log(\tilde{R}_1)}{\rm Re}\left[e^{i\theta} \frac{w'(R e^{i\theta})}{w(R e^{i\theta})}\right] + o(\epsilon^2) \equiv \epsilon f_\theta + o(\epsilon^2)
\end{equation}
where we identify the function $f_\theta$ \rev{referred} to in the main text.
}

\bibliographystyle{apsrev4-2}
\bibliography{main}

@article{Condamin:2008,
	doi = {10.1073/pnas.0712158105},
	author = {Condamin, S. and Tejedor, V. and Voituriez, R. and Benichou, O. and Klafter, J.},
	journal = {Proceedings of the National Academy of Sciences},
	journal1 = {Proceedings of the National Academy of Sciences},
	pages = {5675},
	title = {Probing microscopic origins of confined subdiffusion by first-passage observables},
	ty = {JOUR},
	url = {https://doi.org/10.1073/pnas.0712158105},
	volume = {105},
	year = {2008}}

@article{Majumdar:2010,
	doi = {10.1103/PhysRevLett.104.020602},
	pages = {020602},
	author = {Majumdar, Satya N. and Rosso, Alberto and Zoia, Andrea},
	date = {2010/01/15/},
	day = {15},
	id = {10.1103/PhysRevLett.104.020602},
	j1 = {PRL},
	journal = {Physical Review Letters},
	journal1 = {Phys. Rev. Lett.},
	m1 = {Copyright (C) 2010 The American Physical Society; Please report any problems to prola{\char64}aps.org},
	month = {01},
	number = {2},
	publisher = {American Physical Society},
	sp = {020602},
	title = {Hitting Probability for Anomalous Diffusion Processes},
	ty = {JOUR},
	url = {https://doi.org/10.1103/PhysRevLett.104.020602},
	volume = {104},
	year = {2010}}

@article{Vezzani:2024aa,
	author = {Vezzani, Alessandro and Burioni, Raffaella},
	date = {2024/05/02/},
	day = {02},
	doi = {10.1103/PhysRevLett.132.187101},
	id = {10.1103/PhysRevLett.132.187101},
	j1 = {PRL},
	journal = {Physical Review Letters},
	journal1 = {Phys. Rev. Lett.},
	month = {05},
	number = {18},
	pages = {187101--},
	publisher = {American Physical Society},
	title = {Fast Rare Events in Exit Times Distributions of Jump Processes},
	url = {https://doi.org/10.1103/PhysRevLett.132.187101},
	volume = {132},
	year = {2024}}

@incollection{grebenkov2024target,
	url = {https://doi.org/10.1007/978-3-031-67802-8_1},
	doi = {10.1007/978-3-031-67802-8_1},
	author = {Grebenkov, Denis and Metzler, Ralf and Oshanin, Gleb},
	booktitle = {Target Search Problems},
	pages = {1--29},
	publisher = {Springer},
	title = {Target Search Problems},
	year = {2024}}

@book{bookSid2014,
	url = {https://doi.org/10.1142/9104},
	doi = {10.1142/9104},
	author = {Metzler, R. and Oshanin, G. and Redner, S.},
	publisher = {World Scientific, Singapore},
	title = {First-Passage Phenomena and Their Applications},
	year = {2014}}

@article{Benichou:2011fk,
	doi = {10.1103/RevModPhys.83.81},
	author = {B{\'e}nichou, O. and Loverdo, C. and Moreau, M. and Voituriez, R.},
	date = {2011/03/28/},
	day = {28},
	id = {10.1103/RevModPhys.83.81},
	j1 = {RMP},
	journal = {Reviews of Modern Physics},
	journal1 = {Rev. Mod. Phys.},
	month = {03},
	number = {1},
	pages = {81--129},
	publisher = {American Physical Society},
	title = {Intermittent search strategies},
	ty = {JOUR},
	url = {https://doi.org/10.1103/RevModPhys.83.81},
	volume = {83},
	year = {2011}}

@article{Benichou:2014fk,
	author = {B{\'e}nichou, O. and Voituriez, R.},
	booktitle = {From first-passage times of random walks in confinement to geometry-controlled kinetics},
	date = {2014/6/30/},
	day = {30},
	doi = {10.1016/j.physrep.2014.02.003},
	isbn = {0370-1573},
	journal = {Physics Reports},
	month = {6},
	number = {4},
	pages = {225--284},
	title = {From first-passage times of random walks in confinement to geometry-controlled kinetics},
	ty = {JOUR},
	url = {https://doi.org/10.1016/j.physrep.2014.02.003},
	volume = {539},
	year = {2014}}

@article{Levernier:2018qf,
	author = {Levernier, N. and B{\'e}nichou, O. and Gu{\'e}rin, T. and Voituriez, R.},
	date = {2018/08/23/},
	day = {23},
	doi = {10.1103/PhysRevE.98.022125},
	id = {10.1103/PhysRevE.98.022125},
	j1 = {PRE},
	journal = {Physical Review E},
	journal1 = {Phys. Rev. E},
	month = {08},
	number = {2},
	pages = {022125--},
	publisher = {American Physical Society},
	title = {Universal first-passage statistics in aging media},
	ty = {JOUR},
	url = {https://doi.org/10.1103/PhysRevE.98.022125},
	volume = {98},
	year = {2018}}

@article{Klinger:2022vc,
	author = {Klinger, J. and Barbier-Chebbah, A. and Voituriez, R. and B{\'e}nichou, O.},
	date = {2022/03/14/},
	day = {14},
	doi = {10.1103/PhysRevE.105.034116},
	id = {10.1103/PhysRevE.105.034116},
	j1 = {PRE},
	journal = {Physical Review E},
	journal1 = {Phys. Rev. E},
	month = {03},
	number = {3},
	pages = {034116--},
	publisher = {American Physical Society},
	title = {Joint statistics of space and time exploration of one-dimensional random walks},
	url = {https://doi.org/10.1103/PhysRevE.105.034116},
	volume = {105},
	year = {2022}}

@article{Anikeenko:2009aa,
	author = {Anikeenko, A. V. and Medvedev, N. N. and Kovalev, M. K. and Melgunov, M. S.},
	date = {2009/06/01},
	doi = {10.1007/s10947-009-0061-8},
	id = {Anikeenko2009},
	isbn = {1573-8779},
	journal = {Journal of Structural Chemistry},
	number = {3},
	pages = {403--410},
	title = {Simulation of gas diffusion in porous layers of varying structure},
	url = {https://doi.org/10.1007/s10947-009-0061-8},
	volume = {50},
	year = {2009}}

@article{Hidalgo-Caballero:2024aa,
	author = {Hidalgo-Caballero, Samuel and Cassinelli, Alvaro and Fort, Emmanuel and Labousse, Matthieu},
	date = {2024/04/26/},
	day = {26},
	doi = {10.1103/PhysRevResearch.6.023103},
	id = {10.1103/PhysRevResearch.6.023103},
	j1 = {PRRESEARCH},
	journal = {Physical Review Research},
	journal1 = {Phys. Rev. Res.},
	month = {04},
	number = {2},
	pages = {023103--},
	publisher = {American Physical Society},
	title = {Frugal random exploration strategy for shape recognition using statistical geometry},
	url = {https://doi.org/10.1103/PhysRevResearch.6.023103},
	volume = {6},
	year = {2024}}

@article{Wright1931,
	author = {Wright, S},
	doi = {10.1093/genetics/16.2.97},
	issn = {0016-6731},
	journal = {Genetics},
	month = {mar},
	number = {2},
	pages = {97--159},
	pmid = {17246615},
	title = {{Evolution in Mendelian Populations.}},
	url = {https://doi.org/10.1093/genetics/16.2.97},
	volume = {16},
	year = {1931}}

@book{Feller2:1971,
	url = {https://www.wiley.com/en-us/An+Introduction+to+Probability+Theory+and+Its+Applications%2C+Volume+2%2C+2nd+Edition-p-9780471257097},
	address = {New York},
	author = {Feller, William},
	description = {Feller volume 2},
	mrclass = {60.00},
	mrnumber = {MR0270403 (42 \#5292)},
	pages = {xxiv+669},
	publisher = {John Wiley \& Sons Inc.},
	series = {Second edition},
	title = {An introduction to probability theory and its applications. {V}ol. { II}.},
	year = {1971}}

@article{Kou:2003,
	url = {https://doi.org/10.1239/aap/1051201658},
	author = {Kou, S. G. and Wang, Hui},
	doi = {10.1239/aap/1051201658},
	issn = {00018678},
	journal = {Advances in Applied Probability},
	number = {2},
	pages = {504--531},
	title = {{First passage times of a jump diffusion process}},
	volume = {35},
	year = {2003}}

@article{Majumdar2010,
	url = {https://doi.org/10.1007/s10955-009-9905-z},
	archiveprefix = {arXiv},
	arxivid = {0912.0631},
	author = {Majumdar, Satya N. and Comtet, Alain and Randon-Furling, Julien},
	doi = {10.1007/s10955-009-9905-z},
	eprint = {0912.0631},
	issn = {00224715},
	journal = {Journal of Statistical Physics},
	number = {6},
	pages = {955--1009},
	title = {{Random Convex Hulls and Extreme Value Statistics}},
	volume = {138},
	year = {2010}}

@article{Kyprianou:2022p,
	author = {Kyprianou, Andreas E. and Pardo, Juan Carlos and Watson, Alexander R.},
	doi = {10.1214/12-AOP790},
	fjournal = {The Annals of Probability},
	issn = {0091-1798},
	journal = {The Annals of Probability},
	mrclass = {60G52 (60G18 60G51)},
	mrnumber = {3161489},
	mrreviewer = {Frank Aurzada},
	number = {1},
	pages = {398--430},
	title = {Hitting distributions of {$\alpha$}-stable processes via path censoring and self-similarity},
	url = {https://doi.org/10.1214/12-AOP790},
	volume = {42},
	year = {2014}}

@article{Pillay:2010,
	doi = {10.1137/090752511},
	author = {Pillay, S. and Ward, M. J. and Peirce, A. and Kolokolnikov, T.},
	date = {2010/01/00/},
	day = {00},
	journal = {Multiscale Modeling \& Simulation},
	journal1 = {Multiscale Modeling \& Simulation},
	journal2 = {Multiscale Model. Simul.},
	month = {01},
	number = {3},
	pages = {803--835},
	publisher = {SIAM},
	title = {An Asymptotic Analysis of the Mean First Passage Time for Narrow Escape Problems: Part I: Two-Dimensional Domains},
	ty = {JOUR},
	url = {https://doi.org/10.1137/090752511},
	volume = {8},
	year = {2010}}

@book{Redner:2001a,
	url = {https://doi.org/10.1017/CBO9780511606014},
	doi = {10.1017/CBO9780511606014},
	author = {S. Redner},
	publisher = {Cambridge University Press, Cambridge, England},
	title = {A Guide to First- Passage Processes},
	year = {2001}}

@article{Blumenthal1961,
	author = {Blumenthal, R. M. and Getoor, R K and Ray, D B},
	doi = {10.1090/S0002-9947-1961-0126885-4},
	issn = {0002-9947},
	journal = {Transactions of the American Mathematical Society},
	number = {3},
	pages = {540--554},
	title = {{On the distribution of first hits for the symmetric stable processes.}},
	url = {https://doi.org/10.1090/S0002-9947-1961-0126885-4},
	volume = {99},
	year = {1961}}

@article{Getoor1961,
	url = {https://doi.org/10.1090/S0002-9947-1961-0137148-5},
	author = {Getoor, R. K.},
	doi = {10.1090/S0002-9947-1961-0137148-5},
	pages = {75--90},
	issn = {00029947},
	journal = {Transactions of the American Mathematical Society},
	number = {1},
	title = {{First Passage Times for Symmetric Stable Processes in Space}},
	volume = {101},
	year = {1961}}

@article{Meyer:2011,
	doi = {10.1103/PhysRevE.83.051116},
	author = {Meyer, B. and Chevalier, C. and Voituriez, R. and B\'enichou, O.},
	date = {2011/05/16/},
	day = {16},
	id = {10.1103/PhysRevE.83.051116},
	j1 = {PRE},
	journal = {Physical Review E},
	journal1 = {Phys. Rev. E},
	number = {5},
	pages = {051116},
	publisher = {American Physical Society},
	sp = {051116},
	title = {Universality classes of first-passage-time distribution in confined media},
	ty = {JOUR},
	u1 = {Copyright (C) 2011 The American Physical Society},
	u2 = {Please report any problems to prola{\char64}aps.org},
	url = {https://doi.org/10.1103/PhysRevE.83.051116},
	volume = {83},
	year = {2011}}

@article{BenichouO.:2010,
	doi = {10.1038/nchem.622},
	author = {B{\'e}nichou, O. and Chevalier, C. and Klafter, J. and Meyer, B. and Voituriez, R.},
	date = {2010/06//print},
	isbn = {1755-4330},
	journal = {Nature Chemistry},
	l3 = {http://www.nature.com/nchem/journal/v2/n6/suppinfo/nchem.622_S1.html},
	m3 = {10.1038/nchem.622},
	month = {06},
	number = {6},
	pages = {472--477},
	publisher = {Nature Publishing Group},
	title = {Geometry-controlled kinetics},
	ty = {JOUR},
	url = {https://doi.org/10.1038/nchem.622},
	volume = {2},
	year = {2010}}

@book{VanKampen:1992,
	url = {https://www.sciencedirect.com/book/9780444529657/stochastic-processes-in-physics-and-chemistry},
	author = {Van Kampen, N.G.},
	publisher = {North-Holland personnal library},
	title = {Stochastic Processes in Physics and Chemistry, Third Edition},
	year = {2007}}

@book{Berg:2004,
	url = {https://doi.org/10.1007/b97370},
	doi = {10.1007/b97370},
	author = {H.C. Berg},
	publisher = {Springer, New York},
	title = {E. Coli in Motion},
	year = {2004}}

@article{Araujo:2021,
	url = {https://doi.org/10.1103/PhysRevE.103.L010101},
	archiveprefix = {arXiv},
	arxivid = {2008.03506},
	author = {Ara{\'{u}}jo, Michelle O. and {De Silans}, Thierry Passerat and Kaiser, Robin},
	doi = {10.1103/PhysRevE.103.L010101},
	eprint = {2008.03506},
	issn = {24700053},
	journal = {Physical Review E},
	number = {1},
	pages = {1--6},
	pmid = {33601531},
	title = {{L{\'{e}}vy flights of photons with infinite mean free path}},
	volume = {103},
	year = {2021}}

@article{Blanco:2003a,
	url = {https://doi.org/10.1209/epl/i2003-00208-x},
	doi = {10.1209/epl/i2003-00208-x},
	author = {Blanco,S. and Fournier,R.},
	id = {0295-5075-61-2-168},
	isbn = {0295-5075},
	journal = {EPL (Europhysics Letters)},
	number = {2},
	pages = {168--173},
	title = {An invariance property of diffusive random walks},
	ty = {JOUR},
	volume = {61},
	year = {2003}}

@article{Baudouin:2014,
	url = {https://doi.org/10.1103/PhysRevE.90.052114},
	author = {Baudouin, Q. and Pierrat, R. and Eloy, A. and Nunes-Pereira, E. J. and Cuniasse, P. A. and Mercadier, N. and Kaiser, R.},
	doi = {10.1103/PhysRevE.90.052114},
	issn = {15502376},
	journal = {Physical Review E},
	number = {5},
	pages = {1--11},
	title = {{Signatures of L{\'{e}}vy flights with annealed disorder}},
	volume = {90},
	year = {2014}}

@article{Romanczuk:2012fk,
	url = {https://doi.org/10.1140/epjst/e2012-01529-y},
	af = {Romanczuk, P. Baer, M. Ebeling, W. Lindner, B. Schimansky-Geier, L.},
	author = {Romanczuk, P. and Bar, M. and Ebeling, W. and Lindner, B. and Schimansky-Geier, L.},
	date = {MAR},
	doi = {10.1140/epjst/e2012-01529-y},
	isi = {WOS:000301985100001},
	issn = {1951-6355},
	journal = {The European Physical Journal Special Topics},
	month = {Mar},
	number = {1},
	oi = {Lindner, Benjamin/0000-0001-5617-127X},
	pages = {1--162},
	publication-type = {J},
	ri = {Lindner, Benjamin/D-5438-2009},
	title = {Active Brownian particles From Individual to Collective Stochastic Dynamics},
	volume = {202},
	year = {2012}}

@article{Solon:2015,
	url = {https://doi.org/10.1140/epjst/e2015-02457-0},
	archiveprefix = {arXiv},
	arxivid = {1504.07391},
	author = {Solon, A. P. and Cates, M. E. and Tailleur, J.},
	doi = {10.1140/epjst/e2015-02457-0},
	eprint = {1504.07391},
	issn = {19516401},
	journal = {The European Physical Journal Special Topics},
	number = {7},
	pages = {1231--1262},
	title = {{Active brownian particles and run-and-tumble particles: A comparative study}},
	volume = {224},
	year = {2015}}

@article{Klinger2022,
	author = {Klinger, J. and Voituriez, R. and B{\'{e}}nichou, O.},
	doi = {10.1103/PhysRevLett.129.140603},
	issn = {0031-9007},
	journal = {Physical Review Letters},
	month = {sep},
	number = {14},
	pages = {140603},
	title = {{Splitting Probabilities of Symmetric Jump Processes}},
	url = {https://doi.org/10.1103/PhysRevLett.129.140603},
	volume = {129},
	year = {2022}}

@book{Gnedenko:1955,
	url = {https://archive.org/details/gnedenko-kolmogorov-limit-distributions-for-sums-of-independent-random-variables},
	author = {Boris Gnedenko and Academician A. N. Kolmogorov},
	title = {Limit Distributions for Sums of Independent Random Variables},
	publisher = {Addison-Wesley, Cambridge, MA},
	year = {1954}}

@misc{Stone:1967,
	url = {http://projecteuclid.org/euclid.bsmsp/1200513469},
	author = {Stone, Charles},
	howpublished = {Proc. 5th {Berkeley} {Sympos}. math. {Statist}. {Probab}., { Univ}. {Calif}. 1965/1966, 2, {Part} 2, 217-224 (1967).},
	title = {On local and ratio limit theorems},
	year = {1967},
	zbl = {0236.60021},
	zbmath = {3374724}}

@article{Ivanov1994,
	url = {https://ui.adsabs.harvard.edu/abs/1994A%26A...286..328I/abstract},
	author = {Ivanov, V. V.},
	issn = {0004-6361},
	journal = {Astronomy and Astrophysics},
	number = {1},
	title = {{Resolvent method: exact solutions of half-space transport problems by elementary means}},
	volume = {286},
	year = {1994}}

@article{Sparre:1954,
	doi = {10.7146/math.scand.a-10407},
	author = {Andersen, E. S.},
	journal = {Mathematica Scandinavica},
	pages = {195-223},
	title = {On the fluctuations of sums of random variables II},
	url = {https://doi.org/10.7146/math.scand.a-10407},
	volume = {2},
	year = {1954}}

@article{Majumdar2017,
	author = {Majumdar, Satya N. and Mounaix, Philippe and Schehr, Gr{\'{e}}gory},
	doi = {10.1088/1751-8121/aa8d28},
	issn = {1751-8113},
	journal = {Journal of Physics A: Mathematical and Theoretical},
	month = {nov},
	number = {46},
	pages = {465002},
	publisher = {IOP Publishing},
	title = {{Survival probability of random walks and L{\'{e}}vy flights on a semi-infinite line}},
	url = {https://doi.org/10.1088/1751-8121/aa8d28},
	volume = {50},
	year = {2017}}

@article{Savo:2017,
	author = {Savo, Romolo and Pierrat, Romain and Najar, Ulysse and Carminati, R{ \'{e}}mi and Rotter, Stefan and Gigan, Sylvain},
	doi = {10.1126/science.aan4054},
	issn = {0036-8075},
	journal = {Science},
	month = {nov},
	number = {6364},
	pages = {765--768},
	title = {{Observation of mean path length invariance in light-scattering media }},
	url = {https://doi.org/10.1126/science.aan4054},
	volume = {358},
	year = {2017}}

@article{Blanco:2006,
	doi = {10.1103/PhysRevLett.97.230604},
	author = {Blanco, Stephane and Fournier, Richard},
	journal = {Physical Review Letters},
	journal1 = {Physical Review Letters},
	journal2 = {Phys. Rev. Lett.},
	number = {23},
	pages = {230604--4},
	publisher = {APS},
	title = {Short-Path Statistics and the Diffusion Approximation},
	ty = {JOUR},
	url = {https://doi.org/10.1103/PhysRevLett.97.230604},
	volume = {97},
	year = {2006}}

@article{Caravenna:2008,
	url = {https://doi.org/10.1214/07-AIHP119},
	archiveprefix = {arXiv},
	arxivid = {arXiv:math/0602306v2},
	author = {Caravenna, Francesco and Chaumont, Lo{\"{i}}c},
	doi = {10.1214/07-AIHP119},
	eprint = {0602306v2},
	issn = {02460203},
	journal = {Annales de l'institut Henri Poincare (B) Probability and Statistics},
	number = {1},
	pages = {170--190},
	primaryclass = {arXiv:math},
	title = {{Invariance principles for random walks conditioned to stay positive}},
	volume = {44},
	year = {2008}}

@article{Tailleur:2008,
	author = {Tailleur, J. and Cates, M. E.},
	doi = {10.1103/PhysRevLett.100.218103},
	issn = {0031-9007},
	journal = {Physical Review Letters},
	month = {may},
	number = {21},
	pages = {218103},
	title = {{Statistical Mechanics of Interacting Run-and-Tumble Bacteria}},
	url = {https://doi.org/10.1103/PhysRevLett.100.218103},
	volume = {100},
	year = {2008}}

@article{Edwards2007,
	author = {Edwards, Andrew M. and Phillips, Richard A. and Watkins, Nicholas W. and Freeman, Mervyn P. and Murphy, Eugene J. and Afanasyev, Vsevolod and Buldyrev, Sergey V. and da Luz, M. G. E. and Raposo, E. P. and Stanley, H. Eugene and Viswanathan, Gandhimohan M.},
	doi = {10.1038/nature06199},
	issn = {0028-0836},
	journal = {Nature},
	month = {oct},
	number = {7165},
	pages = {1044--1048},
	title = {{Revisiting L{\'{e}}vy flight search patterns of wandering albatrosses, bumblebees and deer}},
	url = {https://doi.org/10.1038/nature06199},
	volume = {449},
	year = {2007}}

@article{Calandre:2014ul,
	url = {https://doi.org/10.1103/PhysRevE.89.012149},
	doi = {10.1103/PhysRevE.89.012149},
	author = {Calandre, T and B{\'e}nichou, O and Grebenkov, D S and Voituriez, R},
	journal = {Physical Review E},
	journal-full = {Physical review. E, Statistical, nonlinear, and soft matter physics},
	month = {Jan},
	number = {1},
	pages = {012149},
	pmid = {24580214},
	pst = {ppublish},
	title = {Splitting probabilities and interfacial territory covered by two-dimensional and three-dimensional surface-mediated diffusion},
	volume = {89},
	year = {2014}}

@article{Condamin:2007eu,
	url = {https://doi.org/10.1103/PhysRevE.75.021111},
	doi = {10.1103/PhysRevE.75.021111},
	address = {Laboratoire de Physique Theorique de la Matiere Condensee (UMR 7600), Case Courrier 121, Universite Paris 6, 4 Place Jussieu, 75255 Paris Cedex, France.},
	au = {Condamin, S and Benichou, O and Moreau, M},
	author = {Condamin, S and Benichou, O and Moreau, M},
	da = {20070315},
	dcom = {20070501},
	dep = {20070213},
	edat = {2007/03/16 09:00},
	issn = {1539-3755 (Print)},
	jid = {101136452},
	journal = {Physical Review E},
	jt = {Physical review. E, Statistical, nonlinear, and soft matter physics},
	language = {eng},
	mhda = {2007/03/16 09:01},
	number = {2 Pt 1},
	own = {NLM},
	pages = {021111},
	phst = {2006/10/03 {$[$}received{$]$}; 2006/11/28 {$[$}revised{$]$}; 2007/02/13 {$[$}aheadofprint{$]$}},
	pl = {United States},
	pmid = {17358317},
	pst = {ppublish},
	pt = {Journal Article},
	pubm = {Print-Electronic},
	so = {Phys Rev E Stat Nonlin Soft Matter Phys. 2007 Feb;75(2 Pt 1):021111. Epub 2007 Feb 13.},
	stat = {PubMed-not-MEDLINE},
	title = {Random walks and Brownian motion: a method of computation for first-passage times and related quantities in confined geometries.},
	volume = {75},
	year = {2007}}

@article{Chen:2009,
	url = {https://doi.org/10.1002/cae.20208},
	author = {Chen, Jeng Tzong and Tsai, Ming Hong and Liu, Chein Shan},
	doi = {10.1002/cae.20208},
	issn = {10613773},
	journal = {Computer Applications in Engineering Education},
	number = {3},
	pages = {314--322},
	title = {{Conformal mapping and bipolar coordinate for eccentric Laplace problems}},
	volume = {17},
	year = {2009}}

@article{kramers_brownian_1940,
	author = {Kramers, H. A.},
	doi = {10.1016/S0031-8914(40)90098-2},
	issn = {0031-8914},
	journal = {Physica},
	month = apr,
	number = {4},
	pages = {284--304},
	title = {Brownian motion in a field of force and the diffusion model of chemical reactions},
	url = {https://doi.org/10.1016/S0031-8914(40)90098-2},
	urldate = {2024-11-23},
	volume = {7},
	year = {1940}}

@article{Condamin2007,
	url = {https://doi.org/10.1038/nature06201},
	author = {Condamin, S. and B{\'e}nichou, O. and Tejedor, V. and Voituriez, R. and Klafter, J.},
	doi = {10.1038/nature06201},
	journal = {Nature},
	pages = {77--80},
	title = {First-passage times in complex scale-invariant media},
	volume = {450},
	year = {2007}}

@article{Condamin:2005,
	author = {Condamin, S. and B\'enichou, O. and Moreau, M.},
	doi = {10.1103/PhysRevLett.95.260601},
	issue = {26},
	journal = {Physical Review Letters},
	month = {Dec},
	numpages = {4},
	pages = {260601},
	publisher = {American Physical Society},
	title = {First-Passage Times for Random Walks in Bounded Domains},
	url = {https://doi.org/10.1103/PhysRevLett.95.260601},
	volume = {95},
	year = {2005}}

@article{Mori2020a,
	url = {https://doi.org/10.1103/PhysRevLett.124.090603},
	archiveprefix = {arXiv},
	arxivid = {2001.01492},
	author = {Mori, Francesco and {Le Doussal}, Pierre and Majumdar, Satya N. and Schehr, Gr{\'{e}}gory},
	doi = {10.1103/PhysRevLett.124.090603},
	eprint = {2001.01492},
	issn = {10797114},
	journal = {Physical Review Letters},
	number = {9},
	pages = {1--6},
	pmid = {32202896},
	title = {{Universal Survival Probability for a d-Dimensional Run-and-Tumble Particle}},
	volume = {124},
	year = {2020}}

@book{RiceBook,
	url = {https://www.sciencedirect.com/bookseries/comprehensive-chemical-kinetics/vol/25/suppl/C},
	author = {Rice, S},
	publisher = {Elsevier},
	title = {Diffusion-Limited Reactions},
	year = {1985}}

@article{jeanson:2005,
	author = {Raphael Jeanson and Colette Rivault and Jean-Louis Deneubourg and Stephane Blanco and Richard Fournier and Christian Jost and Guy Theraulaz},
	doi = {10.1016/j.anbehav.2004.02.009},
	issn = {0003-3472},
	journal = {Animal Behaviour},
	number = {1},
	pages = {169-180},
	title = {Self-organized aggregation in cockroaches},
	url = {https://doi.org/10.1016/j.anbehav.2004.02.009},
	volume = {69},
	year = {2005}}

@article{Arrhenius:1889,
	author = {Svante Arrhenius},
	title = {{\"U}ber die Dissociationsw{\"a}rme und den Einfluss der Temperatur auf den Dissociationsgrad der Elektrolyte},
	doi = {10.1515/zpch-1889-0408},
	journal = {Zeitschrift f{\"u}r Physikalische Chemie},
	lastchecked = {2025-02-03},
	number = {1},
	pages = {96--116},
	url = {https://doi.org/10.1515/zpch-1889-0408},
	volume = {4U},
	year = {1889}}

@article{RandonFurling:2009,
	author = {Randon-Furling, Julien and Majumdar, Satya N. and Comtet, Alain},
	doi = {10.1103/PhysRevLett.103.140602},
	issue = {14},
	journal = {Physical Review Letters},
	month = {Sep},
	numpages = {4},
	pages = {140602},
	publisher = {American Physical Society},
	title = {Convex Hull of $N$ Planar Brownian Motions: Exact Results and an Application to Ecology},
	url = {https://doi.org/10.1103/PhysRevLett.103.140602},
	volume = {103},
	year = {2009}}

@article{chevalier_first-passage_2011,
	url = {https://doi.org/10.1088/1751-8113/44/2/025002},
	author = {Chevalier, C. and B{\'e}nichou, O. and Meyer, B. and Voituriez, R.},
	doi = {10.1088/1751-8113/44/2/025002},
	issn = {17518113},
	journal = {Journal of Physics A: Mathematical and Theoretical},
	number = {2},
	title = {First-passage quantities of {Brownian} motion in a bounded domain with multiple targets: {A} unified approach},
	pages = {025002},
	volume = {44},
	year = {2011}}

@article{cheviakov_optimizing_2011,
	author = {Cheviakov, A. F. and Ward, M. J.},
	doi = {10.1016/j.mcm.2010.02.025},
	issn = {0895-7177},
	journal = {Mathematical and Computer Modelling},
	number = {7},
	pages = {1394--1409},
	title = {Optimizing the principal eigenvalue of the {Laplacian} in a sphere with interior traps},
	url = {https://doi.org/10.1016/j.mcm.2010.02.025},
	volume = {53},
	year = {2011}}

@article{Skorokhod:1961,
	url = {https://doi.org/10.1137/1106035},
	author = {Skorokhod, A. V.},
	journal = {Theory of Probability \& Its Applications},
	number = {3},
	pages = {264--274},
	title = {Stochastic Equations for Diffusion Processes in a Bounded Region},
	volume = {6},
	year = {1961},
	doi = {10.1137/1106035}}

@book{Whitt:2002,
	url = {https://doi.org/10.1007/b97479},
	doi = {10.1007/b97479},
	author = {Whitt, Ward},
	publisher = {Springer-Verlag, New York},
	series = {Springer Series in Operations Research},
	title = {Stochastic-Process Limits: An Introduction to Stochastic-Process Limits and Their Application to Queues},
	year = {2002}}

@article{Chechkin:2003,
	url = {https://doi.org/10.1088/0305-4470/36/41/L01},
	author = {Chechkin, A. V. and Metzler, R. and Klafter, J. and Gonchar, V. Y. and Tanatarov, L. V.},
	journal = {Journal of Physics A: Mathematical and General},
	number = {41},
	pages = {L537},
	title = {First passage and arrival time densities for {L}\'evy flights and the failure of the method of images},
	volume = {36},
	year = {2003},
	doi = {10.1088/0305-4470/36/41/L01}}

@article{Garbaczewski:2019,
	url = {https://doi.org/10.1103/PhysRevE.99.042126},
	author = {Garbaczewski, Piotr and Stephanovich, Vladimir},
	journal = {Physical Review E},
	number = {4},
	pages = {042126},
	title = {Fractional {L}aplacians in bounded domains: Killed, reflected, censored, and taboo {L}\'evy flights},
	volume = {99},
	year = {2019},
	doi = {10.1103/PhysRevE.99.042126}}

@article{Kwasnicki:2012,
	url = {https://doi.org/10.1016/j.jfa.2011.12.004},
	author = {Kwa\'snicki, Mateusz},
	journal = {Journal of Functional Analysis},
	number = {5},
	pages = {2379--2402},
	title = {Eigenvalues of the fractional {L}aplace operator in the interval},
	volume = {262},
	year = {2012},
	doi = {10.1016/j.jfa.2011.12.004}}
\end{document}